\documentclass{amsart}
\usepackage{amssymb}
\usepackage{amsmath}

\usepackage{a4}
\begin{document}
\newtheorem{theorem}{Theorem}[section]
\newtheorem{lemma}[theorem]{Lemma}
\newtheorem{remark}[theorem]{Remark}
\newtheorem{definition}[theorem]{Definition}
\newtheorem{corollary}[theorem]{Corollary}
\newtheorem{example}[theorem]{Example}
\newtheorem{assumption}[theorem]{Assumption}
\newtheorem{Rem}[theorem]{Remark} \newtheorem{sats}[theorem]{Theorem}
\newtheorem{prop}[theorem]{Proposition}
\newtheorem{lem}[theorem]{Lemma} \newtheorem{kor}[theorem]{Corollary}
\newcommand{\banm}{\begin{anm}}
\newcommand{\eanm}{\end{anm}}
\newcommand{\rank}{{\rm Im}}
\newcommand{\ind}{{\rm ind}}
\newcommand{\ord}{{\rm ord}}
\newcommand{\comp}{{\rm comp}}
\newcommand{\coker}{{\rm coker}}
\newcommand{\diag}{{\rm diag}}
\newcommand{\diam}{{\rm diam}}
\newcommand{\col}{{\rm col}}
\newcommand{\loc}{{\rm loc}}
\newcommand{\grad}{{\rm grad}}
\newcommand{\Dom}{{\rm Dom}}
\newcommand{\const}{{\rm const}}
\newcommand{\mes}{{\rm mes}}
\def\beq{\begin{eqnarray}}
\def\eeq{\end{eqnarray}}
\newcommand{\dr}{\frac{\partial}{\partial r}}
\newcommand{\nn}{\nonumber}
\newcommand{\nats}{\mbox{${\rm I\!N }$}}
\newcommand{\cal}{\mathcal}
\def\qedbox{\hbox{$\rlap{$\sqcap$}\sqcup$}}
\newcommand{\W}{{\mathaccent"7017 W}} \newcommand{\V}{{\mathaccent"7017 V}} \newcommand{\dist}{{\rm dist}} \font\pbglie=eufm10 \def\CD{C_{\mathcal{D}}} \def\CR{C_{\mathcal{R}}} \def\CDR{C_{\mathcal{D,R}}} \def\DD{\text{\pbglie D}} \makeatletter
  \renewcommand{\theequation}{%
   \thesection.\alph{equation}}
  \@addtoreset{equation}{section}
 \makeatother
\def\BB{\mathcal{B}}
\title[Zaremba]
{Heat Content Asymptotics for Riemannian manifolds with Zaremba boundary conditions}
\author{M. van den Berg, P. Gilkey, K.
Kirsten, and V. A. Kozlov}
\begin{address}{MvdB: Department of Mathematics, University of
Bristol, University Walk, Bristol,\newline\phantom{...a}BS8 1TW, U.K.}\end{address}
\begin{email}{M.vandenBerg@bris.ac.uk}\end{email}
\begin{address}{PG: Mathematics Department, University of Oregon, Eugene, OR 97403, USA}\end{address}
\begin{email}{gilkey@darkwing.uoregon.edu}\end{email}
\begin{address}{KK: Department of Mathematics, Baylor University \\ Waco, TX 76798, USA}\end{address} \begin{email}{Klaus\_Kirsten@baylor.edu}\end{email}
\begin{address}{VK: Mathematics Department, Link\"oping University, SE-581 83 Link\" oping, Sweden}\end{address}
\begin{email}{vlkoz@mai.liu.se}\end{email}
\begin{abstract} The existence of a full asymptotic expansion for the heat content asymptotics of
an operator of Laplace type with classical Zaremba boundary conditions  on a smooth manifold  is established. The first three
coefficients in this
asymptotic expansion are determined in terms of geometric invariants; partial information is obtained about the fourth
coefficient.
\end{abstract}
\maketitle
\section{Introduction}
Let $(M,g)$ be a smooth compact $m$-dimensional manifold with smooth boundary $\partial M$ and let
$V$ be a smooth vector bundle over $M$. Let
$$D=-(a^{ij}\operatorname{Id}\cdot\partial_i\partial_j+b^k\partial_k+c)$$
be a smooth second order operator over $M$ with scalar leading
symbol; we adopt the Einstein convention and sum over repeated
indices. We assume the matrix $\{a^{ij}\}$ is positive definite
and use the inverse matrix $g_{ij}$ to define a Riemannian metric
on $M$.

We can write the operator $D$ invariantly as follows. There is a
unique connection $\nabla$ on $V$ and a unique endomorphism $E$ of
$V$ so that $$D=D(\nabla,E)=-(a^{ij}\nabla_i\nabla_j+E)\,.$$
The connection $1$ form and endomorphism $E$ are given in terms of the derivatives of the total symbol of $D$ and the Christoffel symbols $\Gamma$ by:
 \begin{equation}\label{eqn-1.a} \begin{array}{l} \omega_i=\frac12g_{ij}b^j+\frac12a^{kl}\Gamma_{kli}\\
E=c-a^{ij}(\partial_i\omega_j+\omega_i\omega_j-\omega_k\Gamma_{ij}{}^k)\,.
   \vphantom{\vrule height 11pt}
\end{array}\end{equation}

The boundary conditions we shall impose are at the heart of the
matter. We assume given a decomposition $\partial M=\CR \cup \CD $
as the union of two closed submanifolds with common smooth
boundary $\CR \cap \CD =\Sigma$. Let $\phi_{;m}$ denote the
covariant derivative of $\phi$ with respect to the inward unit
normal on $\partial M$. Let $S$ be an auxiliary endomorphism of
$V|_{\CR }$. We take Robin boundary conditions on $\CR $ and
Dirichlet boundary conditions on $\CD $ arising from the boundary
operator: \begin{equation}\label{eqn-1.b}
\BB\phi:=(\phi_{;m}+S\phi)|_{\{\CR-\Sigma\}}\oplus \phi|_{\CD}.
\end{equation} We refer to Seeley \cite{Se02,Se03} for a more general formalism; see also related work of Avramidi
\cite{Av03}, Dowker \cite{Doa,Dob}, and Jakobson et al. \cite{JLNP}.

Let $e^{-tD_\BB}$ be the fundamental solution of the heat
equation; $u=e^{-tD_\BB}\phi$ is then characterized by the
equations: \begin{equation}\label{eqn-1.c} (\partial_t+D)u=0,\
u(x;0)=\phi(x),\ \BB u(x,t)=0\text{ for }t>0. \end{equation} The
equality $u(x;0)=\phi(x)$ is to be taken in the $L^2$ sense where
$\phi\in C^\infty(M;V)$ is a smooth section to $V$ which gives the
initial temperature distribution. Let $\phi^*\in C^\infty(M;V^*)$
be a smooth section to the dual bundle $V^*$ which gives the
specific heat of the manifold. We denote the natural pairing
between $V$ and $V^*$ by $\langle\cdot,\cdot\rangle$. Let $dx$,
$dx^\prime$, and $dz$ be the Riemannian measures on $M$, on
$\partial M$, and on $\Sigma$, respectively. We define the {\it
total heat energy content} of the manifold by setting:
$$\beta(\phi,\phi^*,D,\BB)(t):=\int_M\langle
e^{-tD_\BB}\phi,\phi^*\rangle dx\,.$$

It is worth putting this in a more classical framework in the
special case where $D=\Delta$ is the Laplacian and $S=0$. Let
$W^{1,2}(M)$ be the closure of $C^\infty(M)$ with respect to the
Sobolev norm
$$||\phi||_1^2=\int_M\{|\nabla\phi|^2+|\phi|^2\}dx.$$
Let $W^{1,2}_{0,C_D}(M)$ be the closure of the set
$$\{\phi\in W^{1,2}(M):\text{supp}(\phi)\cap C_D=\emptyset\}\,.$$
Thus, for example, $W_{0,\emptyset}^{1,2}(M)=W^{1,2}(M)$. Let $W_0^{1,2}(M)=W_{0,\partial M}^{1,2}(M)$. For $\lambda>0$, let
$$N(M,C_D,\lambda)=\sup(\dim E_\lambda)$$
where the supremum is taken over all subspaces $E_\lambda\subset W_{0,C_D}^{1,2}(M)$ such that
$$||\nabla\phi||_{L^2(M)}<\lambda||\phi||_{L^2(M)},\quad\forall\phi\in E_\lambda\,.$$
Then $N(M,\emptyset,\lambda)$ is the spectral counting function for the Neumann Laplacian on $M$, $N(M,\partial M,\lambda)$ is
the spectral counting function for the Dirichlet Laplacian on $M$, and $N(M,C_D,\lambda)$ is the spectral counting function for
the Laplacian acting in
$L^2(M)$ with Dirichlet conditions on $C_D$ and Neumann conditions on $\partial M-C_D$.

It is well known, see for example McKean and Singer \cite{MS67}, that since $M$ is compact and $\partial M$ is smooth,
$N(M,\emptyset,\lambda)$ is finite. By the variational principle,
$$N(M,\partial M,\lambda)\le N(M,C_D,\lambda)\le N(M,\emptyset,\lambda);$$
consequently $N(M,C_D,\lambda)$ is finite and counts the number of eigenvalues less than $\lambda$ with the Zaremba boundary
condition defined by $C_D$. Let $\lambda_1\le\lambda_2\le...$ be the eigenvalues counted by $N(M,C_D,\cdot)$ and let $\{\phi_i\}$
be a corresponding orthonormal basis of eigenfunctions in $L^2(M)$. If $\phi$ and $\phi^*$ are smooth, we can express the heat
content in terms of the Fourier coefficients:
\begin{equation}\label{eqn-1.dx}
\beta(\phi,\phi^*,D,\BB)(t)=\sum_ie^{-t\lambda_i}\langle\phi,\phi_i\rangle_{L^2(M)}\langle\phi^*,\phi_i\rangle_{L^2(M)}\,.
\end{equation}
By Parseval's identity, the series converges for $\phi=\phi^*=1$ and for all $t>0$:
\begin{eqnarray*}\beta(1,1,D,\mathcal{B})(t)&=&\sum_ie^{-t\lambda_i}\left\{\langle1,\phi_i\rangle_{L^2(M)}\right\}^2\\
&\le&e^{-t\lambda_1}\sum_i\langle1,\phi_i\rangle_{L^2(M)}^2=\textstyle{vol}(M)e^{-t\lambda_1}\,.
\end{eqnarray*}
It now follows that the series in Equation (\ref{eqn-1.dx}) converges for all smooth $\phi$ and $\phi^*$ and for all $t>0$.

We return to the general setting. Adopt the notation established
above.

\begin{theorem}\label{thm-1.1}
Let $\phi\in C^\infty(M;V)$ and let $\phi^*\in C^\infty(M;V^*)$.
There exists a complete asymptotic expansion
$\beta(\phi,\phi^*,D,\BB)(t)\sim\sum_{n\ge0}\beta_n(\phi,\phi^*,D,\BB)t^{n/2}$
where the $\beta_n$ are locally computable in terms of integrals
over $M$, $\CD $,  $\CR$, and $\Sigma$. \end{theorem}

We use the dual connection $\tilde\nabla$ to covariantly
differentiate sections of $V^*$. Let the dual endomorphism $\tilde
S$ on $V^*$ define the dual boundary condition $\tilde\BB$ for the
dual operator $\tilde D$ on $\CR$. Near the boundary, we choose a
local orthonormal frame $\{e_i\}$ for the tangent bundle of $M$ so
that $e_m$ is the inward unit geodesic normal vector field; on
$\Sigma$, we assume $e_{m-1}$ is the inward unit normal of
$\Sigma\subset\CD$. Let indices $a,b$ range from $1$ through $m-1$
and index the induced orthonormal frame for the tangent bundle of
the boundary; let indices $u,v$ range from $1$ through $m-2$ and
index the induced orthonormal frame for $\Sigma$. Let
$L_{ab}:=\Gamma_{abm}$ and $\tilde L_{uv}:=\Gamma_{uv(m-1)}$ be the
components of the second fundamental forms of $\partial M\subset
M$ and $\Sigma\subset\CD$, respectively. Let $R$ be the Riemann
curvature tensor with the sign convention that $R_{1221}=+1$ for
the standard sphere in $\mathbb{R}^3$.

\begin{theorem}\label{thm-1.2} There exist universal constants $c_i$ so that:
\begin{enumerate} \item $\beta_0(\phi,\phi^*,D,\BB)=\int_M\langle\phi,
\phi^*\rangle dx$. \item $\beta_1(\phi,\phi^*,D,\BB)=
    -\frac2{\sqrt\pi}\int_{\CD }\langle\phi,\phi^*\rangle dx^\prime$.
    \item $\beta_2(\phi,\phi^*,D,\BB)=-\int_M\langle D\phi,\phi^*\rangle dx
+\int_{\CR
}\{\langle\phi_{;m}+S\phi,\phi^*\rangle\}dx^\prime$
\smallbreak\quad$+\int_{\CD}
\{\frac12L_{aa}\langle\phi,\phi^*\rangle-\langle\phi,\phi^*_{;m}\rangle\}dx^\prime
+c_0\int_\Sigma\langle\phi,\phi^*\rangle dz$.
\item $\beta_3(\phi,\phi^*,D,\BB)=\frac4{3\sqrt\pi}\int_{\CR }\langle\phi_{;m}
+S\phi,\phi^*_{;m}+\tilde S\phi^*\rangle
dx^\prime$\smallbreak\quad$ -\frac2{\sqrt\pi}\int_{\CD
}\{
\frac23\langle\phi_{;mm},\phi^*\rangle+\frac23\langle\phi,\phi^*_{;mm}\rangle
-\langle\phi_{:a},\phi^*_{:a}\rangle +\langle
E\phi,\phi^*\rangle$
\smallbreak\quad$-\frac23L_{aa}\langle\phi,\phi^*\rangle_{;m}
+\langle(\frac1{12}L_{aa}L_{bb}-\frac16L_{ab}L_{ab}+\frac16R_{amam})
\phi,\phi^*\rangle\}dx^\prime$\smallbreak\quad
$+\int_{\Sigma}\{
\langle(c_1L_{m-1,m-1}+c_2L_{uu}+c_3\tilde
L_{uu}+c_4S)\phi,\phi^*\rangle$ \smallbreak\qquad\quad$
+c_5\langle\phi,\phi^*\rangle_{;m-1}+c_6\langle\phi,\phi^*\rangle_{;m}
\}dz$.
\item We have $c_0=-\frac12$, $c_3 = \frac 1 {2\sqrt \pi}$, $c_5 = \frac 1 {2\sqrt
\pi}$, $c_6 = - \frac 2 {3\sqrt \pi}$.
\item We  have $c_2-\frac12c_4=\frac 2 {3\sqrt \pi}$.
\end{enumerate} \end{theorem}

\begin{remark}\rm Our methods did not yield $c_1$. They also did not permit us to complete the computation of $\{c_2,c_4\}$.
\end{remark}

Theorem \ref{thm-1.2} follows if $\Sigma$ is empty from results of
\cite{BeDeGi93,BeGi94}; the new feature here is the additional
integrands over $\Sigma$ present in $\beta_2$ and $\beta_3$ and
the partial information we have obtained concerning these terms.

Here is a brief guide to the paper.
In Section \ref{sect-3}, we use invariance theory to establish
Assertions (1-4) of Theorem \ref{thm-1.2}. In Section
\ref{sect-4}, we use product formulas and make a special case
computation to show
\begin{equation}\label{eqn-1.d}
c_0=-\frac12,\quad
\frac1{\sqrt\pi}-c_3-c_5=0,\quad\text{and}\quad c_2-\frac12c_4+c_6=0\,.
\end{equation}
In Section \ref{sect-4a}
we use special cases on the half-plane to complete the proof by showing
\begin{equation}\label{eqn-1.e}
c_3 = \frac 1 {2\sqrt \pi},\quad c_5 = \frac 1 {2\sqrt
\pi},\quad\text{and}\quad c_6 = - \frac 2 {3\sqrt \pi}\,.
\end{equation}

The remainder of the paper is devoted to the proof of Theorem
\ref{thm-1.1} using results of \cite{Ko88,KM1,KM2}; we believe the
methods of Seeley \cite{Se02,Se03} could also be used. We switch focus
completely at this stage. Instead of working invariantly and
globally in the context of Riemannian manifolds using methods of
invariance theory, we work locally in Euclidean space in a system
of local coordinates. In Section \ref{sect-5}, we introduce the function spaces which we shall need. In Section \ref{sect-6}, we
discuss various `model' problems and in Section \ref{sect-7}, we
formulate a basic theorem on asymptotics. We obtain a number of
estimates and conclude the proof of Theorem \ref{thm-1.1} in Section \ref{sect-8} by establishing a slightly more general result
(see Theorem
\ref{thm-8.2}).

\section{Universal expressions for the invariants $\beta_n^\Sigma$}
\label{sect-3} Let $\beta_n^\Sigma$ denote the additional
invariant defined by integration over $\Sigma$. Dimensional
analysis then shows that $\beta_n^\Sigma$ can be computed by
integrating invariants which are homogeneous of weight $n-2$ in
the jets of the symbol; see the discussion in
\cite{BeDeGi93,BeGi94} on this point where a similar analysis was
performed which studied the interior invariants and the boundary
invariants for Dirichlet or Neumann problems. Thus trivially,
$\beta_0^\Sigma=\beta_1^\Sigma=0$, and therefore the first
interesting contribution arises at the $\beta_2$ level; this must
be a constant multiple of $\langle\phi,\phi^*\rangle$. Assertions
(1-3) of Theorem \ref{thm-1.2} now follow.

To study the form of the additional boundary integral over
$\Sigma$ appearing in $\beta_3$, we investigate the local geometry
near $\Sigma$. Fix a point $z_0\in\Sigma$. Let
$z=(z_1,...,z_{m-2})$ be local coordinates on $\Sigma$ so
$$g(\partial_u^z,\partial_v^z)(z_0)=\delta_{uv}\quad\text{and}\quad
  \partial_u^zg(\partial_v^z,\partial_w^z)(z_0)=0\quad\text{for}\quad 1\le u,v,w\le m-2.$$
  By considering the geodesic flow from $\Sigma\subset \CD $, we introduce coordinates
  $$(z,y_1)\rightarrow\exp_z\{y_1e_{m-1}(z)\}$$
so that $y_1$ is the signed geodesic distance from $\Sigma$ to
$\partial M$; $\CD $ corresponds to $y_1\ge0$ and $\CR $
corresponds to $y_1<0$. The metric then satisfies:
$$g(\partial_u^z,\partial_1^y)=0\quad\text{and}\quad
g(\partial_1^y,\partial_1^y)=1.$$ Now use the geodesic flow of
$\partial M$ in $M$ to introduce coordinates
$$(z,y_1,y_2)\rightarrow\exp_{(z,y_1)}\{y_2e_m(z,y_1)\}$$ where
$y_2$ is the geodesic distance to $\partial M$. We then have
\begin{eqnarray*} g(\partial_u^z,\partial_2^y)=g(\partial_1^y,\partial_2^y)=0\quad\text{and}\quad
g(\partial_2^y,\partial_2^y)=1\,.
\end{eqnarray*}
The only non-zero derivatives of the metric at $z_0$ are then given by the second fundamental forms:
\begin{eqnarray*} &&\tilde L_{uv}(z_0)=-\frac12\partial_1^yg(\partial_u^z,\partial_v^z)(z_0),\\
&&L_{uv}(z_0):=-\frac12\partial_2^yg(\partial_u^z,\partial_v^z)(z_0),\quad\text{and}\\
&&L_{m-1,m-1}(z_0):=-\frac12\partial_2^yg(\partial_1^y,\partial_1^y)(z_0)\,.
\end{eqnarray*}

The structure group is the orthogonal group $O(m-2)$ and we apply
H. Weyl's Theorem \cite{We46} on the invariants of the orthogonal
group. Assertion (2) now follows by writing down a basis for the
set of invariants of weight $1$ and applying the symmetry
$$\beta_n(\phi,\phi^*,D,\BB)=\beta_n(\phi^*,\phi,\tilde
D,\tilde\BB)$$ where $\tilde D$ and $\tilde\BB$ are the dual
operator and dual boundary condition on the dual bundle $V^*$,
respectively. The usual product and addition formulas then show
the constants are dimension free and universal. This completes the
proof of Theorem \ref{thm-1.2} (4).

\section{Relations among the universal coefficients}\label{sect-4}

The universal coefficient $c_0$ of Theorem \ref{thm-1.2} (3) can be determined by a special case calculation.
Let $M_+$ be a compact convex subset of $\mathbb{R}^2$ with non-empty interior and smooth boundary $\partial M_+$.
We suppose that $\partial M_+$ contains a closed line segment $\Lambda$ of positive length.
Let $M_-$ be the reflection of $M_+$ with respect to the line determined by $\Lambda$.
We assume that $\Lambda=\partial M_+\cap\partial M_-$. We set $N=M_+\cup M_-$.

 Let $\Delta=-\partial_1^2-\partial_2^2$ be the usual flat Laplacian.
 Let
$$\begin{array}{ll} \CR(M_\pm):=\Lambda,&\CD (M_\pm):=\partial M_\pm-\Lambda,\\
 \CR(N):=\emptyset,&\CD (N):=\partial N=\CD(M_+)\cup\CD(M_-).\vphantom{\vrule height 12pt}\end{array}$$
We take $\phi=\phi^*=1$ and $S=0$ to define $u_\pm$ on $M_\pm$ and
$u_N$ on $N$. Let
\begin{equation}\label{eqn-4.a} c(\gamma):=4\int_0^\infty\frac{\sinh((\pi-\gamma)s)}
   {\sinh(\pi s)\cdot\cosh(\gamma s)}ds\,.
\end{equation}

The manifold $N$ has two cusps of angle $2\pi$ at
$\partial\Lambda$. The results of \cite{BeGa94,BeSr90} can be used
to compute $\beta_n$, while Theorem \ref{thm-1.2} (3) can be used
to see: \begin{eqnarray}\label{eqn-4.b}
\beta_N(\phi,\phi^*,\Delta,\BB)(t)&=&\int_Ndx
 -\sqrt t\left\{\frac2{\sqrt\pi}\int_{\partial N}dx^\prime\right\}\\
&+& t\left\{\frac12\int_{\partial N}L_{aa}dx^\prime+2c(2\pi)\right\}+O(t^\frac32),\nonumber\\
\beta_{M_+}(\phi,\phi^*,\Delta,\BB)(t)&=&\int_{M_+}dx
  -\sqrt t\left\{\frac{2}{\sqrt\pi}\int_{\partial M_+}dx^\prime\right\}\nonumber\\
  &+& t\left\{\frac12\int_{\CD (M_+)}L_{aa}dx^\prime+2c_0\right\}+O(t^\frac32)\,.\nonumber
\end{eqnarray}
We have by symmetry that $u_N(\phi,\phi^*,\Delta,\BB)=u_\pm(\phi,\phi^*,\Delta,\BB)$
if $x\in M_\pm$ and $t>0$ and trivially the normal derivative of $N$ vanishes on $\Lambda$. Thus:
\begin{equation}\label{eqn-4.c} \begin{array}{l} \beta_N(\phi,\phi^*,\Delta,\BB)(t)=\beta_{M_+}
(\phi,\phi^*,\Delta,\BB)(t)+\beta_{M_-}(\phi,\phi^*,\Delta,\BB)(t)\\
\qquad\qquad\qquad\qquad=2\beta_{M_+}(\phi,\phi^*,\Delta,\BB)(t).\vphantom{\vrule
height 12pt}\end{array} \end{equation} Since $\partial N=\CD
(M_+)\cup \CD (M_-)$, we may use equations (\ref{eqn-4.b}) and
(\ref{eqn-4.c}) to see that $4c_0=2c(2\pi)$. By Equation
(\ref{eqn-4.a}), $c(2\pi)=-1$. We may therefore conclude
$$c_0=-\textstyle\frac12\,.$$

We use warped product formulae to obtain the two relationships between the coefficients given in Equation (\ref{eqn-1.d}). Let
$M_1:=[0,1]\times S^1$ be the cylinder with the usual parameters $(r,\theta)$. Set: $$\begin{array}{ll}
ds_1^2:=dr^2+d\theta^2,&\CD:=\{0,1\}\times[0,\pi],\\
\CR:=\{0,1\}\times[\pi,2\pi],&
  \Sigma:=\{0,1\}\times\{0,\pi\},\vphantom{\vrule height 11pt}\\ \Delta_1:=-\partial_r^2-\partial_\theta^2,&S_1:=0,
  \vphantom{\vrule height 11pt}\\ \phi_1:=1,&\phi_1^*=1\,.\vphantom{\vrule height 11pt} \end{array}$$
  Since $\Sigma$ is discrete, $dz$ is counting measure. Since all the structures are flat,
$$\beta_n(\phi_1,\phi^*_1,\Delta_1,\BB_1)=0\quad\text{for}\quad n\ge3.$$

Let $\varepsilon$ be a small real parameter. Let $M_2:=M_1\times
S^1$ and let $\Theta$ be the usual periodic parameter on the
second circle. Let $f=f(r,\theta)$ be a smooth warping function
and define $$\begin{array}{ll} ds^2_2=ds^2_1+e^{2\varepsilon
f(r,\theta)}d\Theta^2,& D_2:=\Delta_1-e^{-2\varepsilon
f(r,\theta)}\partial_\Theta^2,\\
\operatorname{dvol}_2=e^{\varepsilon f}drd\theta d\Theta\,.
\vphantom{\vrule height 11pt} \end{array}$$ We take the warping
function to vanish identically near $r=1$ and focus attention on
$r=0$. Note that $D_2$ is not self-adjoint. We let $\BB_2=\BB_1$
induce the same boundary conditions; we must adjust $S_2$
appropriately to once again take pure Neumann boundary conditions
on $\CR \times S^1$ as the connection induced by $D_2$ having a
non-trivial connection $1$ form. We let $\phi_2=1$, but we set
$\phi^*_2=e^{-\varepsilon f(r,\theta)}$ to compensate for the
change in the volume element. Set $u_2=u_1$. We verify that
$u_2=e^{-t\Delta_{2,\BB_2}}\phi_2$ by computing:
\begin{eqnarray*} &&(\partial_t+D_2)u_2=(\partial_t+\Delta_1)u_1=0,\nonumber\\
&&u_2(r,\theta,\Theta;0)=u_1(r,\theta;0)=1,
\\
&&\BB_2u_2=0.\nonumber\end{eqnarray*} Consequently we may compute:
\begin{eqnarray}\nonumber &&\beta(1,\phi^*_2,\Delta_2,\BB_2)(t)=
\int_{M_2}u_2(r,\theta,\Theta;t)
   \phi^*_2(r,\theta,\Theta)e^{\varepsilon f(r,\theta)}drd\theta d\Theta\\
   &&\nonumber\qquad=2\pi\int_{M_1}u_1(r,\theta,\Theta;t)drd\theta
=2\pi\beta(1,1,\Delta_1,\BB_1)(t),\quad\text{so}\\
&&\beta_3(1,\phi^*_2,D_2,\BB_2)=2\pi\beta_3(1,1,\Delta_1,\BB_1)=0\,.\label{eqn-2.d}
\end{eqnarray}

First take $f(r,\theta)=f(\theta)$ to be independent of the radial parameter near $r=0$.
We use equation (\ref{eqn-1.a}) to see: $$\begin{array}{ll} \omega_r=\omega_\Theta=0,& \omega_\theta=-
\textstyle\frac\varepsilon2f_\theta,\\
\nabla_\theta\phi_2=(\partial_\theta+\omega_\theta)\phi_2=
-\textstyle\frac\varepsilon2f_\theta,&
\tilde\nabla_\theta\phi^*_2=(\partial_\theta-\omega_\theta)\phi^*_2
   =-\textstyle\frac\varepsilon2f_\theta,\\
-\phi_{2:a}\phi^*_{2:a}=-\textstyle\frac{\varepsilon^2}4f_\theta
f_\theta e^{-\varepsilon f(\theta)}, &E=
\textstyle\frac\varepsilon2f_{\theta\theta}+\frac{\varepsilon^2}4f_\theta
f_\theta e^{-\varepsilon f(\theta)}\,.
\end{array}$$
Consequently we have
\begin{equation}\label{eqn-2.e}
\beta_3^{\CD
}(\phi_2,\phi^*_2,D_2,\BB)=-\frac2{\sqrt\pi}\int_{\CD
} \frac\varepsilon2 f_{\theta\theta}d\theta d\Theta=
\frac\varepsilon{\sqrt\pi}\int_\Sigma f_\theta
d\Theta\,. \end{equation}
Note that $c_3\tilde
L_{uu}\langle\phi_2,\phi^*_2\rangle=-c_3f_\theta\varepsilon
e^{-\varepsilon f}$ and
$c_5\langle\phi_2,\phi^*_2\rangle_{;m-1}=-c_5f_{\theta}\varepsilon
e^{-\varepsilon f}$. Because $$
\beta_3^\Sigma(\phi_2,\phi^*_2,D_2,\BB)=-\varepsilon\int_\Sigma
(c_3+c_5)f_\theta d\Theta,$$ we have the desired
relationship $$\textstyle\frac1{\sqrt\pi}-c_3-c_5=0\,.$$

We now take $f(r,\theta)=f(r)$ where
$$f(0)=0,\ \partial_rf(0)=1,\text{ and
}\partial_r^kf(0)=0\text{ for }k>1.$$
Since $\BB\phi_2=0$ on $\CR
$, only $\CD $ and $\Sigma$ are relevant. We follow the discussion
in Section 3 of \cite{BeGi94} to show $\beta_3^{\CD}=0$. We have
$$\begin{array}{lll}
\phi=1,&\phi^*=e^{-\varepsilon f(r)},&
  \Gamma_{\Theta\Theta r}=-\textstyle\varepsilon e^{\varepsilon f},\\ \omega_r=-\textstyle\frac\varepsilon2,&
\tilde\omega_r=\textstyle\frac\varepsilon2,&S_2=\textstyle\frac12\varepsilon\,.\vphantom{\vrule
height 11pt} \end{array}$$ Consequently, we may compute on $\CD$
that: \begin{eqnarray*}
&&\textstyle\frac23\langle\phi_{;mm},\phi^*\rangle+\textstyle\frac23\langle\phi,\phi^*_{;mm}\rangle
=\frac{\varepsilon^2}3,\\
&&E=-\omega_r\omega_r+\omega_r\Gamma_{\theta\theta
r}=\textstyle\frac{\varepsilon^2}4,\\
&&-\textstyle\frac23L_{aa}\langle\phi,\phi^*\rangle_{;m}=
-\frac{2\varepsilon^2}3,\\
&&\textstyle(\frac1{12}L_{aa}L_{bb}-\frac16L_{ab}L_{ab}+\frac16R_{amam})
=(\frac1{12}-\frac16+\frac16)\varepsilon^2,\\
&&\beta_3^{\CD
}=\textstyle(\frac{4}{12}+\frac{3}{12}-\frac8{12}+\frac1{12})\varepsilon^2=0.
\end{eqnarray*}
This implies that
\begin{eqnarray*} 0&=&\beta_3^\Sigma(1,\phi^*_2,\Delta_2,\BB_2)\\
&=&\int_\Sigma\{\langle(c_2L_{uu}+c_4S)\phi_2,\phi^*_2\rangle+c_6\langle\phi_2,\phi_2^*\rangle_{;m}\}dz\\
&=&(-c_2+\frac12c_4-c_6)\varepsilon\operatorname{vol}(\Sigma)\,.
\end{eqnarray*}
This establishes Equation (\ref{eqn-1.d}) by showing
that $$c_2-\textstyle\frac12c_4+c_6=0\,.$$

\section{An example on the half-plane}\label{sect-4a}
We now consider an example on the half-plane, where the classical Zaremba boundary
value problem that we are considering has a simple spectral resolution. The two-dimensional Laplacian is given in polar
coordinates by
\beq
\Delta =
\frac 1 r \frac{\partial}{\partial r} \left( r \dr \right) + \frac 1 {r^2}
\frac{\partial ^2}{\partial \varphi ^2} .\nn\eeq

We let $\phi=0$ define the Dirichlet component and $\phi=\pi$ define the Neumann boundary component. The spectral
resolution is then given by:
\beq \psi _{\lambda, k}
(\varphi , r) = \sqrt{\frac 2 \pi } \sin \left( \varphi [k+1/2]
\right) J_{k+1/2} (\lambda r) , \quad k\in\nats_0.\label{kk1a}
\eeq We can can use this spectral resolution to write down
the Fourier decomposition of the heat content where we assume a
suitable decay of $\phi$ and $\phi^*$ at infinity to ensure this is well defined:\begin{eqnarray*}&& \beta (\phi ,
\phi^* ,{\mathcal D}, {\mathcal B} ) (t) = \sum_{k=0} ^\infty
\int_0^\infty d\lambda \,\, \lambda e^{-t \lambda ^2}
\gamma_{k,\lambda} (\phi ) \gamma_{k,\lambda} (\phi^*) \quad\text{where}\\
&& \gamma_{k,\lambda} (f) = \int_0^\pi d\varphi
\int_0^\infty dr \,\, r f(\varphi , r) \psi_{\lambda ,k} (\varphi,
r) \,.\end{eqnarray*}

We first perform the $\lambda$ integration using \cite{grad65},
Equation 6.633, \beq \int_0^\infty dx \,\, x e^{-t x^2} J_p
(\alpha x) J_p (\beta x) = \frac 1 {2t} e^{-\frac{\alpha^2 +
\beta^2} {4t}} I_p \left( \frac{ \alpha \beta} {2t} \right)
.\nn\eeq This leads to the following representation of the heat
content, \beq \beta (\phi , \phi^* , {\cal D}, {\cal B} ) (t) &=&
\frac 1 {\pi t} \sum_{k=0} ^\infty \int_0^\pi d\varphi
\int_0^\infty dr \, r \int_0^\pi d\varphi ' \int_0^\infty dr' \,
r' \phi (\varphi, r)
\phi^* (\varphi ' , r' ) \nn\\
& & \quad \sin \left( \varphi [ k+1/2] \right) \sin \left(
\varphi ' [k+1/2] \right) e^{ -\frac{ r^2 +r'^2} {4t} } I_{k+1/2}
\left(\frac{ r r'} {2t} \right) .\label{kk1}\eeq We will choose
suitable angular parts for the localizing functions $\phi ,
\phi^*$ to ensure that the angular integrals can be obtained in closed
form. As we will see, an arbitrary $r$-dependence can be dealt
with.

The basis for the forthcoming calculation is the integral
representation of the Bessel function $I_{k+1/2}$ where $k$ is an integer (we refer to \cite{grad65}, Equation
8.431.5 for details):
\beq I_{k+1/2} (z) &=& \frac 1 \pi \int_0^\pi d\theta \,
e^{z \cos \theta} \cos \left(
[k+1/2] \theta \right) \nn\\
& &-\frac{\sin \left( [ k+1/2] \pi \right) }\pi \int_0^\infty
d\tau
\,\, e^{-z \cosh \tau - \left( k+\frac 1 2 \right) \tau } \nn\\
&=& \frac 1 \pi \int_0^\pi d\theta \, e^{z \cos \theta} \cos
\left(
[k+1/2] \theta \right) \nn\\
& &-\frac 1 \pi (-1) ^k  \int_0^\infty d\tau \,\, e^{-z \cosh \tau
- \left( k+\frac 1 2 \right) \tau } . \label{kk3} \eeq
\begin{remark}\rm If we had studied pure Dirichlet or pure Neumann boundary conditions, then the relevant
Bessel functions would be indexed by an integer rather than by the half integer $k+1/2$. The second term in the representation
given by Equation (\ref{kk3}) would be absent in such a case.\end{remark}

Substituting the identity
\beq \exp \left\{ -
\frac{r^2 + {r'} ^2} {4t} + \frac{ r r'}{2t} \cos \theta \right\}
=\exp \left\{ - \frac{(r- r') ^2} {4t} + \frac{ r r'}{2t} (\cos
\theta -1) \right\} \label{kk2}\eeq
into (\ref{kk1}) and using a saddle point argument, we can verify that the first term
in (\ref{kk3}) is `responsible' for producing the volume and
boundary contributions.  We will show that the second term is `responsible'
for the contributions concentrated on $\Sigma$.
As we are interested in the contributions concentrated on $\Sigma$, which we will denote by $\beta^\Sigma$,
we only study the second term in (\ref{kk3}).

 We give another
derivation of the identity $c_0 = - 1/2$ to illustrate the general idea behind the calculation. We assume $\phi$ and $\phi^*$ to
have the product form \beq \phi = \Omega_1 (\varphi ) R_1 (r) , \quad \quad \phi^*
= \Omega _2 (\varphi ) R_2 (r) .\nn\eeq

We first study a constant
angular part $\Omega _i(\varphi ) =1$, $i=1,2$, and perform the
angular integrations, \beq \int_0^\pi d\varphi \sin \left( \varphi
\left[ k + \frac 1 2 \right] \right) = \frac 1 {k+1/2} .\nn\eeq
This yields the identity \beq \beta^\Sigma (\phi , \phi^*, {\cal D}, {\cal B}
) (t) &=& - \frac 1 {\pi^2 t} \int_0^\infty dr r \int_0 ^\infty
dr' r' R_1 (r) R_2 (r') e^{-\frac{r^2 + {r'}^2} {4t} } \nn\\
 & &\quad \int_0^\infty
 d\tau e^{-\frac{ r r'} {2t} \cosh \tau } \sum_{k=0} ^\infty
 (-1)^k \frac{ e^{-(k+1/2) \tau }}{(k+1/2)^2} .\nn\eeq
Thus only $r\sim 0$ and $r' \sim 0$
contribute to the asymptotic small $t$ expansion of the heat
content.

We substitute $y=r/\sqrt t$, $y' = r' /\sqrt t$, and
expand around $r=0$, $r' =0$. To leading order this produces \beq
\beta^\Sigma (\phi , \phi^*, {\cal D}, {\cal B} ) (t) &\sim& -
\frac t {\pi^2} R_1 (0) R_2 (0) \int_0^\infty d\tau \sum_{k=0}
^\infty (-1)^k \frac{ e^{-(k+1/2) \tau}}{(k+1/2)^2} \nn\\ & &
\quad \int_0^\infty dy \int_0^\infty dy' y y' e^{-\frac{ y^2 +
{y'}^2} 4 - \frac { yy'} 2 \cosh \tau} .\nn\eeq

The constant $c_0$ is determined by this triple integral and the sum
over $k$. We perform the $y'$-integral using \cite{grad65},
Equation 3.322.2; this involves a complementary error function
which, together with \cite{grad65}, Equation 6.286.1, yields \beq \beta^\Sigma (\phi , \phi^*, {\cal D}, {\cal
B} ) (t) &\sim& - \frac t {\pi^2} R_1 (0) R_2 (0) \sum_{k=0} ^\infty
\frac{(-1)^k}{(k+1/2)^2} \nn\\
& &\quad\int_0^\infty d\tau  e^{-(k+1/2) \tau} \left\{ - \frac 4
{\sinh ^2 \tau } + \frac{ 4\tau \cosh \tau}{\sinh^3 \tau}
\right\}.\label{kk4} \eeq The integration of the single terms in
(\ref{kk4}) is not possible as is seen from the $\tau \to 0$
behaviour. In order to use \cite{grad65}, Equation 3.541.1, \beq
\int_0^\infty d\tau e^{-\mu \tau} \sinh ^\alpha (\beta \tau) =
\frac 1 {2^{\alpha +1} \beta} B \left( \frac{ \mu }{2\beta} -
\frac \alpha 2 , \alpha +1 \right) ,\label{kk5}\eeq with the
beta-function $B(x,y)$, we need to introduce a regularizing factor
$\sinh^\nu \tau$ in (\ref{kk4}), integrate the single terms and
perform the limit $\nu \to 0^+ $ at the end of the calculation. We
obtain
\begin{eqnarray*}
&&\beta^\Sigma (\phi , \phi^*, {\cal D}, {\cal
B} )
(t)  \\
 &\sim& - \frac t {\pi^2} R_1 (0) R_2 (0) \sum_{k=0}^\infty\frac{(-1)^k}{(k+1/2)^2} \left[
-2 \left( k+\frac 3 2 \right) + \left( k+\frac 1 2 \right) ^2 \psi
' \left( \frac 1 2 \left[ k + \frac 1 2 \right] \right) \right]
\end{eqnarray*}
with the psi-function $\psi (x) = (d/dx)\ln\Gamma (x)$.
The first term can be summed with the aid of the Hurwitz zeta
function, \beq \sum_{k=0} ^\infty (-1)^k \frac{ k + 3/2} {
(k+1/2)^2}= \frac \pi 2 + 4 C,\nn\eeq $C$ being the Catalan
constant. The summation over the derivatives of the
$\psi$-function is performed using \cite{grad65}, Equation
8.363.8, \beq \psi ' (x) = \zeta _H (2;x). \nn\eeq We use this to
write \beq \lefteqn{\sum _{k=0} ^\infty (-1)^k \psi ' \left( \frac
1 2 \left[ k+\frac 1 2 \right] \right) =\sum _{k=0} ^\infty (-1)^k
\sum_{l=0} ^\infty \frac 1 { \left( l + \frac 1 2 \left[ k+\frac 1
2 \right]
\right) ^2 } }\nn\\
&=& \lim_{\nu \to 0^+} \sum_{k=0}^\infty (-1)^k \sum_{l=0} ^\infty
\frac 1 { \left( l + \frac 1 2 \left[ k+\frac 1 2 \right]
\right) ^{2+\nu} } \nn\\
&= & \lim_{\nu \to 0^+} \sum_{l=0} ^\infty \sum_{k=0} ^\infty
\left\{ \frac 1 {
(l+k+1/4)^{2+\nu}} - \frac 1 {(l+k+3/4)^{2+\nu}}\right\} \nn\\
&=& \lim_{\nu \to 0^+} \sum_{m=0} ^\infty \sum_{l=0} ^m \left\{
\frac 1 { (m+1/4)^{2+\nu}}
- \frac 1 {(m+3/4)^{2+\nu}} \right\} \nn\\
&=& \lim_{\nu \to 0^+} \sum_{m=0}^\infty \left\{ \frac{
m+1}{(m+1/4)^{2+\nu}} - \frac {m+1}{(m+3/4)^{2+{\nu}}} \right\} \nn\\
&=& \lim_{\nu \to 0^+} \left( \zeta _H (1+\nu ; 1/4) - \zeta_H
(1+\nu ; 3/4) \right) + \frac 3 4 \zeta_H (2;1/4) - \frac 1 4
\zeta (2;3/4) \nn\\
&=& - \psi \left( \frac 1 4 \right) + \psi \left( \frac 3 4
\right) + \frac 3 4 \zeta _H \left( 2; 1/4\right) - \frac 1 4
\zeta _H \left( 2; 3/4 \right) = \pi + 8 C + \frac {\pi^2} 2.
\nn\eeq This allows us to conclude \beq\beta^\Sigma (\phi ,
\phi^*, {\cal D}, {\cal B} ) (t) &\sim& -\frac 1 2 t R_ 1 (0) R_2
(0) , \nn\eeq which gives us another derivation of the result that $c_0 = -1/2$. Since we have chosen a
constant angular dependence, we do not have $<\phi , \phi ^* >
_{;m}$ terms. Also, given the localizing functions are assumed to
be $C^\infty (M; V)$, we need $R_i (r) = R_i (-r)$, which implies
$(\partial /
\partial r) R_i (r) |_{r=0} =0$; so we do not have $<\phi , \phi^*
> _{; m-1}$ terms either and we need not consider other terms in the
asymptotic expansion for this example.

In order to obtain information about constants
$c_5$ and $c_6$, we need to study nontrivial angular dependences.
The constant $c_6$ is studied by looking at \beq \Omega _1
(\varphi ) =1 , \quad \quad \Omega _2 (\varphi ) = \sin \varphi\,.
\nn\eeq Assuming $R_2 (r) = r R_2 ' (0) +
{\cal O} (r^2)$ as $r\to 0$, we have that $\phi ^* _{;m}|_{r=0} = R_2 '
(0)$.

As in the previous calculation, we
start by observing that \beq \int_0^\pi d\varphi \,\, \sin \left(
\varphi \left[ k+\frac 1 2 \right]\right) \sin \varphi =
(-1)^{k+1} \left( \frac 1 {2k-1} - \frac 1 {2k+3}\right).\nn\eeq
This allows us to write the leading term of $\beta^\Sigma$ as
$t\to 0$ in the form
 \beq &&\beta^\Sigma (\phi , \phi^*, {\cal D}, {\cal
B} ) (t)\nonumber\\ &\sim&  \frac {2t^{3/2}} {\pi^2} R_1 (0) R_2 '(0)
\sum_{k=0}^\infty \left\{
\frac 1 {(2k-1)(2k+1)} - \frac 1 {(2k+1) (2k+3)} \right\} \nn\\
& &\quad \int_0^\infty d\tau  e^{-\left(k+\frac 1 2\right)
\tau}\int_0^\infty dy \int_0^\infty dy' y {y'}^2 e^{-\frac{ y^2 +
{y'}^2} 4 - \frac { yy'} 2 \cosh \tau} .\nn\eeq To
proceed as before with \cite{grad65}, Equations 3.322.2 and
6.286.1, we observe \beq I&:=& \int_0^\infty { dy^\prime}
\int_0^\infty dy \,\, y {y'}^2
e^{-\frac{y^2+{y'}^2} 4 - \frac{yy'} 2 \cosh\tau} \nn\\
&=& -\frac 2 {\sinh \tau} \frac d {d\tau} \int_0^\infty dy' y'
e^{-\frac{{y'}^2} 4 } \int_0^\infty dy e^{-\frac{y^2} 4 - \frac{ y
y'} 2 \cosh\tau} . \nn\eeq The $y$-integral is evaluated using
\cite{grad65}, Equation 3.322.2, the resulting $y'$-integral with
\cite{grad65}, Equation 6.286.1. This shows that \beq I = -
\frac{2\sqrt{\pi}} {\sinh \tau} \frac d {d\tau} \frac 1 {\cosh ^2
\tau} {_2F_1} \left( 1 , \frac 3 2 ; 2; \tanh ^2 \tau \right) ,
\nn\eeq with the hypergeometric function $_2F_1 (a,b;c;x)$. For
the particular parameters involved, the hypergeometric function is
\beq _2F_1 \left( 1 , \frac 3 2 ; 2 ; x\right) = - \frac { 2 (-1 +
\sqrt{ 1-x} )} { x \sqrt{ 1-x}} , \nn\eeq which for $x=\tanh ^2
\tau$ yields the identity \beq _2F_1 \left( 1 , \frac 3 2 ; 2 ; \tanh^2 \tau
\right) = 2 \frac{ \cosh ^2 \tau }{\sinh ^2 \tau} (-1 + \cosh \tau
) . \nn\eeq This shows that \beq I = - 4 \sqrt \pi \left\{
\frac { 2\cosh \tau} { \sinh^4 \tau} - \frac 2 { \sinh^4 \tau} -
\frac 1 {\sinh^2 \tau}\right\} = \frac{ \sqrt \pi}{\cosh ^4 \left(
\frac \tau 2 \right)} , \nn\eeq and consequently we have that \beq
&&\beta^\Sigma (\phi , \phi^*, {\cal D}, {\cal B} ) (t)\nn\\ &\sim&
\frac{ 2 t^{\frac 3 2 }}{\pi^{\frac 3 2 }} R_1 (0) R_2 ' (0)
\sum_{k=0}^\infty \left\{ \frac 1 {(2k-1) ( 2k+1)} -
\frac 1 { (2k+1) ( 2k+3)}\right\} \nn\\
& &\times\int_0^\infty d\tau \,\, e^{-\left( k+\frac 1 2 \right)
\tau} \frac 1 {\cosh ^4 \left( \frac \tau 2 \right)}
.\label{kk6}\eeq We evaluate the integral using
\cite{grad65}, Equation 3.541.8. With $\mu = k+1/2$ and with the
standard notation \cite{grad65} \beq \beta (x) = \frac 1 2 \left[
\Psi \left( \frac{x+1} 2 \right) - \Psi \left( \frac x 2 \right)
\right] = \sum_{k=0}^\infty \frac{ (-1)^k}{x+k}\,.\nn\eeq This shows that
\beq \int\limits_0^\infty \frac {e^{-\mu \tau}}{\cosh ^4 \left(
\frac \tau 2 \right)}{ d\tau} = \frac 2 3 \mu \left\{ 1 + 2 (\mu ^2 -1) [
\beta (\mu +1) - \beta (\mu )] \right\}.\nn\eeq We note that \beq
\frac 1 {(2 k-1) (2 k+1)} - \frac 1 {(2k+1) (2k+3)} = \frac 1 {
(\mu -1) \mu (\mu +1)} , \nn\eeq which allows us to write
\begin{eqnarray}
&&\beta^\Sigma (\phi , \phi^*, {\cal D}, {\cal B} ) (t)\nonumber\\&& \sim \frac{
2 t^{\frac 3 2 }}{3\pi^{\frac 3 2 }} R_1 (0) R_2 ' (0)
\sum_{k=0}^\infty \left\{ \frac 1 {(\mu -1) (\mu +1) } + 2 [\beta
(\mu +1) - \beta (\mu) ] \right\} . \label{kk7}\end{eqnarray}

Our final
task is the evaluation of the sum over $k$. To this end, we note
that \beq&& \sum_{k=0}^\infty \frac 1 { \left( k -\frac 1 2
\right) \left( k + \frac 3 2 \right) }=0 ,\quad\text{and}\quad\nn\\
&&\sum_{k=0}^\infty (\beta (k+3/2) - \beta (k+1/2) ) = -\beta
(1/2) = - \frac \pi 2 .\nn\eeq We may then conclude that \beq \beta^\Sigma (\phi ,
\phi^*, {\cal D}, {\cal B} ) (t) \sim -\frac 2 {3\sqrt \pi} R_1
(0) R_2 ' (0) t^{3/2}\,.\nn\eeq This shows, as desired, that \beq c_6 =
-\frac 2 {3\sqrt \pi} .\nn\eeq

In order to determine $c_5$, we choose\beq \Omega _1(\varphi ) = 1 \quad \text{and}\quad
\Omega _2 (\varphi ) = \cos \varphi\,.\nn\eeq
We again suppose that $R_2 (r) = r R_2 ' (0) + {\cal O} (r^2)$. Then as $r\to
0$, \beq \phi ^* _{; m-1} |_{r=0}= R_2 ' (0) .\nn\eeq The relevant
angular part integration is \beq \int_0^\pi d \varphi \sin \left(
\varphi \left[ k + \frac 1 2 \right] \right) \cos \varphi = \frac
1 {2k-1} + \frac 1 {2k+3} .\nn\eeq The $y$ and $y'$ integration,
as well as the resulting $\tau$-integration are the same as
before, and the equation corresponding to (\ref{kk7}) for this
example is \beq \beta^\Sigma (\phi , \phi^*, {\cal D}, {\cal B} )
(t) \sim -
\frac{  t^{\frac 3 2 }}{3\pi^{\frac 3 2 }} R_1 (0) R_2 '(0)\nn\\
& &\hspace{-3.0cm}\times \sum_{k=0}^\infty  (-1)^k \left\{ \frac 1
{\mu -1} + \frac 1 {\mu +1} + 4 \mu ( \beta (\mu +1) - \beta (\mu
)) \right\} . \label{kk8}\eeq The remaining sum over $k$ may be
performed with the help of the identities \beq &&\sum_{k=0}^\infty
\frac{(-1)^k} { k-\frac 1 2 } = -\frac 1 2
(\pi +4) ,\quad\sum_{k=0}^\infty  \frac{(-1)^k}{ k+\frac 3 2 } = \frac 1 2
(4-\pi),\quad\text{and}\nn\\
&&\sum_{k+0}^\infty (-1)^k (k+1/2) (\beta (k+3/2) - \beta (k+1/2))
= -\frac \pi 8\,.\nn\eeq We add these relations to conclude that
\beq \beta^\Sigma (\phi , \phi^*, {\cal D}, {\cal B} ) (t) &\sim&
\frac 1 {2 \sqrt \pi} R_1 (0) R_2 ' (0) t^{3/2} \,.\nn\eeq This shows that
\beq c_5 = \frac 1 {2\sqrt \pi} .\nn\eeq This establishes Equation (\ref{eqn-1.e}) and thereby completes the proof of
Theorem \ref{thm-1.2}.

\section{Function Spaces}\label{sect-5}

The analysis in question is local so we shall suppose $M$ is an
open domain in $\mathbb{R}^m$ with compact closure and with smooth
boundary $\partial M$. For the sake of simplicity, we shall assume
that the vector bundle in question is trivial; the analysis is
similar in the bundle valued case. We write the Robin boundary
operator in local coordinates in the form $$ Ru=\sum
a^{ij}(x)\sigma_i\partial_{x_j}+d(x),
$$
where $\sigma =(\sigma_1,\ldots ,\sigma _m)$ is the  outward unit
normal to $\partial M$  and $d$ is a smooth function on $\partial
M$.

Let $\nu$ be the distance to $\partial M$ and let $x'$ denote a
point on $\partial M$. We introduce coordinates
$(x^\prime,\nu)\rightarrow x^\prime+\nu \sigma$ on a collared
neighborhood of the boundary to express
\begin{equation}\label{Rep1} D(x,\partial
_x)=-a(x^\prime)\partial_\nu^2+\nu b(x^\prime,\nu
)\partial_\nu^2+L_1(x^\prime,\nu ,\partial_{x^\prime})\partial_\nu
+L_2(x^\prime,\nu ,\partial_{x^\prime})\,. \end{equation} In
Equation (\ref{Rep1}), $a$ is a smooth positive function on
$\partial M$, $b$ is smooth in a neighborhood of the boundary, and
$L_1$ and $L_2$ are differential operators in $x^\prime$ of orders $1$, $2$,
respectively, with smooth coefficients near $\partial M$.

Near $\Sigma $, $M$ is diffeomorphic to $\Sigma\times
B_+(\varepsilon )$ for some $\varepsilon>0$, where
$$B_+(\varepsilon ):=\{
y=(y_1,y_2)\, :\, y_2\geq 0,\, y_1^2+y_2^2<\varepsilon ^2\}\,.$$
We may choose $y_2$ to be the normal parameter $\nu$ and use
coordinates $(z,y)$ near $\Sigma $. With these normalizations,
$$\CD=\{(z,y):y_2=0,y_1\ge0\}\quad\text{and}\quad \CR=\{(z,y):y_2=0,y_1<0\}\,.$$ We can express the operator $D$ as: \begin{equation}\label{Rep2}\begin{array}{ll}
D(x,\partial_x)=&-L(z,\partial_y)\textstyle
+A_2(z,y,\partial_z)\\
&+\textstyle
\sum_{i=1,2}(y_iB_{i2}(z,y,\partial_{y})+A_{i1}(z,y,\partial_z)\partial_{y_i})\,.
\vphantom{\vrule height 11pt}
\end{array}\end{equation}
In this formulation, $A_{i1}$ and $B_{i2}$, $A_2$ are differential
operators with smooth coefficients of orders $1$ and $2$,
respectively. Furthermore, $L$ can be expressed as
\begin{equation}\label{LLa}
L(z,\partial_y)=\textstyle\sum_{i,j=1,2}A^{ij}(z)\partial_{y_i}\partial_{y_j}\,
, \end{equation} where $A^{ij}$ are smooth real valued functions
on $\Sigma$ such that the matrix $\{ A^{ij}\}$ is symmetric
positive definite and \begin{equation}\label{LLb} L(z,\xi
)\leq\varkappa^{-1}_0|\xi |^2\;\;\;\mbox{for all $z\in\Sigma$ and
$\xi\in\mathbb{R}^{m-2}$} \end{equation} with some positive
$\varkappa_0$. The boundary operator $R$ can be represented in
these coordinates as \begin{equation}\label{Rep3}
R(x^\prime,\partial_x)=R_0(z,\partial_y)
+y_1B_{1}(z,y_1,\partial_{y})+A_1(z,y,\partial_z)\, ,
\end{equation} where $B_1$ and $A_1$ are differential operators
with respect to $y$ and $z$ of orders $1$ with smooth
coefficients, and \begin{equation}\label{Rep4}
R_0(z,\partial_y)=\textstyle\sum_{j=1,2}A^{j2}(z)\partial_{y_j}\,
. \end{equation}

We blowup $\Sigma$ and introduce polar coordinates near $\Sigma$
by setting:
$$(z,\rho,\theta)\rightarrow (z,y)=(z, \rho \cos\theta, \rho \sin \theta).$$ Note that $\theta=0$ defines $\CD $ while $\theta=\pi$ defines $\CR $. 

Let $C^\infty(M_\Sigma )$ be the class of smooth functions on
$C^\infty(M-\Sigma)$ which extend smoothly to the blowup. In what
follows we shall suppose that a function $\rho $, which defined
only locally, has been extended smoothly as a $C^\infty(M_\Sigma)$
function which is positive outside the original
neighborhood.

\begin{definition}\label{defn5.1}
\rm For $\kappa>0$, let ${\mathcal E}(\kappa )$ be the set of
functions $U\in C^\infty ([0,\infty ))$ which satisfy
estimates of the form $$ |\partial^k_\tau U(\tau )|\leq
C_{k,\kappa '}\exp (-\kappa '\tau^2)\quad\text{for}\quad
k=0,1,2,...\quad\text{and}\quad\kappa^\prime\in(0,\kappa)\,.
$$
The optimal constants $C_{k,\kappa^\prime}$ define semi-norms
$$p_{k,\kappa '}(U)
:=\inf C_{k,\kappa '}$$ giving a Frechet space topology on
${\mathcal E}(\kappa )$. We use this topology to define subspaces of $\mathcal{E}(\kappa)$
of smooth functions on $\CD$ and on $\CR$: $$ {\mathcal E}(\CD,\kappa )=C^\infty
(\CD,{\mathcal E}(\kappa ))\quad\text{and}\quad {\mathcal
E}(\CR,\kappa )=C^\infty (\CR,{\mathcal E}(\kappa ))\,. $$
\end{definition}

Introduce the halfspace $\mathbb{R}^2_+=\{ y=(y_1,y_2):y_2\geq
0\}$ with polar coordinates $(\rho,\theta)$ for
$\rho\in[0,\infty)$ and $\theta\in[0,\pi]$.
\begin{definition}\label{defn-5.2}
\rm Let $\mathcal{U}$ be a smooth function on
$\mathbb{R}^2_+-\{0\}$. We say that ${\mathcal U}\in\Lambda
^\mu_\kappa $ if\begin{enumerate} \item $
|\partial_\rho^k\partial_\theta ^j{\mathcal U}(y)|\leq C_{k,j}\rho
^{\mu -k}\;\;\;\mbox{for $\rho\leq 1$}$ and for all $k$, $j$.
\item For large values of $\rho$, the function ${\mathcal U}$ admits an asymptotic expansion
\begin{equation}\label{eqn-5.g}
 {\mathcal U}(y)\sim \sum_{j=0}^\infty \rho^{-j}\left\{\frac{v_j(\theta )}
{\rho}+\sum_{\pm }U_j^{\pm }(y_2)\chi (\textstyle\frac{y_2}\rho )
\right\}\, ,
\end{equation}
where $v_j\in C^\infty ([0,\pi ])$, where $\chi$ is a
smooth cutoff function on $\mathbb{R}_+$ which equals $1$ for small
$\tau $ and $0$ for large $\tau $, and where $U_j^{\pm }\in
{\mathcal E}(\varkappa )$ with $``+"$ corresponding to $y_1>0$ and
$``-"$ corresponding to $y_1<0$. \end{enumerate}\end{definition}
The asymptotic expansion in Definition \ref{defn-5.2} is to be
understood in the following sense. For any $N=1,2,\ldots $ and for
any multi-indices $\alpha$ and $\gamma$ with $|\gamma |\leq |
\alpha |$, we have a constant $C$, which is independent of
$\theta$ and of $\rho$, so that:
\begin{equation}\label{eqn-5.h}\left |y^\gamma\partial
_y^\alpha\left(U(\rho ,\theta )-\sum_{j=0}^{N-1}
\rho^{-j}\left\{\frac{v_j(\theta )}{\rho}+\sum_{\pm }U_j^{\pm
}(y_2)\chi ({\textstyle\frac{y_2}\rho} )
\right\}\right)\right|\leq C\rho ^{-N}\,,\rho\ge1\,.
\end{equation}
One verifies
that the class given in Definition \ref{defn-5.2} is independent
of the particular $\chi$ chosen.

\begin{definition}\label{defn-5.3}
\rm Let $\Lambda ^\mu_\kappa (\Sigma )=C^\infty (\Sigma ;\Lambda
^\mu_\kappa )$ be the set of all functions ${\mathcal U}={\mathcal
U}(z,y)$ from $C^\infty (\Sigma\times (\mathbb{R}^2_+-\{0\}))$
belonging to $\Lambda ^\mu_\kappa $ for every $z\in\Sigma$.  The coefficients $v_j$ and $U_j^{\pm }$ in the
asymptotic expansion (\ref{eqn-5.g}) and in the inequality (\ref{eqn-5.h}) may depend on
$z\in\Sigma$. We assume that these coefficients belong to $C^\infty (\Sigma ,{\mathcal E}(\kappa
))$ and that (\ref{eqn-5.g}) and (\ref{eqn-5.h}) can be differentiated
with respect to $z$. \end{definition}

\begin{definition}\label{defn-5.4}\rm Let $R_\mu$ be the set of smooth functions on $\mathbb{R}_+$ so that:
 \begin{enumerate}
\item $|\partial_\rho^k{\mathcal V}(\rho )|\leq C_k\rho ^{\mu -k}$ for $\rho \leq 1$, \item
$ {\mathcal V}(\rho )\sim \sum _{j=0}^\infty a_j\rho^{-j}$ as $\rho\rightarrow\infty$. \end{enumerate}
\end{definition}
\begin{remark}\rm Condition (2) of Definition \ref{defn-5.4} means
that we have estimates $$
|\partial_\rho^k ( {\mathcal V}(\rho )-\sum _{j=0}^N a_j\rho^{-j} ) |\leq C_{N,k}\rho^{-N-k-1} $$ for all $N$ and $k$ and for
$\rho\ge1$.\end{remark}
We put $${\mathcal R}_\mu
(\Sigma ):=C^\infty (\Sigma ,R_\mu )\,.$$

\section{Model problems}\label{sect-6}
We shall first consider boundary value problems on a half-line and
then subsequently consider boundary value problems on the
half-plane. We begin with the {\it Dirichlet problem}:
\begin{equation}\label{Eq5}\textstyle (\partial_t-\partial^2_\nu
)t^{k/2}U\Big (\frac{\nu }{2\sqrt{t}}\Big )=t^{k/2-1}F\Big
(\frac{\nu }{2\sqrt{t}}\Big ),\;\;\; U(0)=G\, ,
\end{equation}
where $\nu \geq 0$, $t>0$ and $k=0,1,\ldots$ The function $U=U(\nu
)$ then satisfies $$ U^{''}(\nu )+2\nu U'(\nu )-2kU(\nu )=-4F(\nu
)\;\;\;\mbox{for $\nu\geq 0$, $U(0)=G$}. $$ The homogeneous
equation for $U$ (with $F=0$) has two solutions $$\textstyle
\psi_k(\nu )=\int_\nu^\infty (s-\nu
)^ke^{-s^2}ds\quad\text{and}\quad \phi _k(\nu
)=e^{-\nu^2}\partial_\nu^ke^{\nu^2}\, . $$ Therefore if $F=0$ the
only solution to (\ref{Eq5}) decaying for large $\nu $ is the
function $$U(\nu )=2b\psi_k(\nu )/\Gamma ((k+1)/2)\, .$$ Let
$G=0$. We impose suitable decay properties on $F$ to ensure the
following integrals converge and set: $$\textstyle U(\nu )=\left\{
\begin{array}{ll}\frac{4}{k!} (\psi_k(\nu )\int_0^\nu
e^{s^2}\phi_k(s)F(s)ds + \phi_k(\nu )\int_\nu^\infty
e^{s^2}\psi_k(s)F(s)ds )&,k\text{ odd },\\ \\ \frac{4}{k!}
(\psi_k(\nu )\int_0^\nu e^{s^2}\phi_k(s)F(s)ds + \phi_k(\nu
)\int_\nu^\infty e^{s^2}\psi_k(s)F(s)ds )\\
\qquad\textstyle-\frac{8\psi_k(\nu )}{\Gamma
((k+1)/2)(k/2)!}\int_0^\infty e^{s^2}\psi_k(s)F(s)ds&,k\text{ even }.
\end{array}\right.$$

\begin{prop}\label{Prop1a} Let $a\in C^\infty (\CD)$ be a positive function, let $F\in {\mathcal E}(\CD,\kappa )$, let $G\in
C^\infty (\CD)$, and let
$\kappa \leq\min_{x^\prime\in \CD} a^{-1}(x^\prime)$. Then there exists $U\in {\mathcal E}(\CD ,\kappa )$ so $$\textstyle
(\partial_t-a\partial^2_\nu )t^{k/2}U(\frac{\nu }{2\sqrt{t}} )=t^{k/2-1}F (\frac{\nu }{2\sqrt{t}} )\quad\text{and}\quad U(0)=G\,.
$$
\end{prop}

\begin{proof} We make the change of variable $\tau =at $ to reduce the problem of Proposition \ref{Prop1a} to that given in Equation (\ref{Eq5}) with $F=F(x^\prime)$ and $G=G(x^\prime)$ for $x^\prime\in \CD$. One can then use the formulae given above to see that $U\in {\mathcal E}(\CD ,\kappa )$ as claimed. \end{proof}

A similar argument can be given to deal with the {Neumann
problem}: \begin{prop} Let $a\in C^\infty (\CR)$ be a positive
function, let $F\in {\mathcal E}(\CR,\kappa )$, let $H\in C^\infty (\CR)$, and let
$\kappa \leq\min_{x^\prime\in \CR} a^{-1}(x^\prime)$.
Then there exists $U\in {\mathcal E}(\CR ,\kappa )$ so
$$\textstyle (\partial_t-a\partial^2_\nu )t^{k/2}U
(x^\prime,\frac{\nu }{2\sqrt{t}} )=t^{k/2-1}F (x^\prime,\frac{\nu
}{2\sqrt{t}} )\quad\text{and}\quad U'(0)=H\,. $$ \end{prop}

Next we study a model problem in the half-space $\mathbb{R}^2_+$:
\begin{eqnarray} \label{K1a} &&\textstyle (\partial_t-\Delta_y)
(t^{k/2}{\mathcal U} (\frac{y}{2\sqrt{t}} ) )=t^{k/2-1}{\mathcal
F} (\frac{y}{2\sqrt{t}} )\;\;\;\mbox{for $y\in \mathbb{R}_+^2$ and
$t>0$,}\\ \label{K1b} &&\textstyle{\mathcal U}(y_1 ,0)={\mathcal
G}(\rho )\;\;\mbox{for $y_1>0$}\;\;\mbox{and}\;\;
\rho\partial_{y_2}U(y_1,0)={\mathcal H}(\rho )\;\;\mbox{for
$y_1<0$,} \end{eqnarray} where $k$ is a nonnegative integer, where
${\mathcal F}\in\Lambda _\kappa^\mu $, and where ${\mathcal G},\,
{\mathcal H}\in R_\mu $. Equation (\ref{K1a}) can be rewritten as
\begin{equation}\label{K2a}\textstyle
 (\partial_\rho^2+\frac{1}{\rho }\partial_\rho
+\frac{1}{\rho^2}\partial_\theta^2 ){\mathcal
U}+(2\rho\partial_\rho +-2k){\mathcal U}=-4{\mathcal
F}\;\;\;\mbox{on $\mathbb{R}_+^2$}\,. \end{equation}

We adopt the notation of Definition \ref{defn-5.4}. We omit
details of the proof of the following Theorem as it is analogous
to the proof given for Proposition 2 \cite{Ko88}.

\begin{sats}\label{TtTa} Let $\kappa\in (0,1)$ and $\mu\in (-1/2,1/2)$. Let ${\mathcal F}\in \Lambda_\kappa^{\mu -2}$ and ${\mathcal G}\, ,{\mathcal H}\in R_\mu$. Then there exists a unique solution to Equations  {\rm (\ref{K1b})} and {\rm (\ref{K2a})} belonging to $ \Lambda_\kappa^\mu $. \end{sats}

Next we consider the case when the operator and the right-hand
sides in Equations (\ref{K1b}) and (\ref{K2a})  depend on a
parameter. Let $R_0$ be the operator of Equation (\ref{Rep4}), let
$L(z,\partial_y)$ be the operator of Equation {\rm (\ref{LLa})},
and let $\varkappa_0$ be the constant of Equation (\ref{LLb}).

\begin{sats}\label{T1d}Let $\kappa\in (0,\varkappa_0]$
and $\mu\in (-1/2,1/2)$. If ${\mathcal F}\in \Lambda_\kappa^{\mu
-2}(\Sigma )$ and if ${\mathcal G}\, ,{\mathcal H}\in R_\mu
(\Sigma )$, then there exists a unique element ${\mathcal U}\in
\Lambda_\kappa^{\mu }(\Sigma )$ such that \begin{eqnarray*}
&&\textstyle(\partial_t-L(z,\partial_y)) (t^{k/2}{\mathcal U}
(z,\frac{y}{2\sqrt{t}} ) )=t^{k/2-1}{\mathcal F}
(z,\frac{y}{2\sqrt{t}} )\;\;\;\mbox{for $y\in \mathbb{R}_+^2$ and
$t>0$},\\ &&{\mathcal U}(z,y_1,0)={\mathcal G}(z,y_1)\;\;\mbox{for
$y_1>0$,}\\ && \rho (R_0(z,\partial_{y}){\mathcal U}))(z,y_1
,0)={\mathcal H}(z,-y_1 )\;\;\mbox{for $y_1<0$}\,. \end{eqnarray*}
\end{sats}

\begin{proof} Let $B=B_{ij}=A^{-1/2}$ where $A=A^{ij}$. Set $Y_k=\sum_{j=1,2}B_{kj}y_j$.
Then the equations given in Theorem \ref{T1d}  become Equations (\ref{K1a}) and (\ref{K1b}) where the right-hand side depends on
the parameter $z\in\Sigma $. The desired result now follows from Theorem \ref{TtTa}. \end{proof}

\section{A Theorem on asymptotics}\label{sect-7}
We can now establish the result from which Theorem \ref{thm-1.1}
will follow. Let $$ \kappa =\min_{x^\prime\in\partial M}\varkappa
(x^\prime) $$ where $\varkappa (x^\prime)$ is the best constant in
the inequality $$ \sum
a^{ij}(x^\prime)\xi_i\xi_j\leq\varkappa^{-1}(x^\prime)|\xi
|^2\quad\text{for}\quad\xi\in\mathbb{R}^m\,.
$$

\begin{sats}\label{TQuQu}
Let $\Phi\in C^\infty (M_\Sigma )$.
 Then the weak solution of Equation {\rm (\ref{eqn-1.c}}) has the asymptotic
representation:
\begin{eqnarray*}
&&u(x,t)\sim \sum _{k=0}^\infty t^{k/2}\Big \{ \Xi\Big (\frac{\rho
}{2\sqrt{t}}\Big )\Big (u_k(x)+\rho\eta (\nu /\rho )U_k^D\Big
(x^\prime,\frac{\nu }{2\sqrt{t}}\Big )\\ &&+\rho\eta (\nu /\rho
)U_k^R\Big (x^\prime,\frac{\nu }{2\sqrt{t}}\Big )\Big )+\zeta
(\rho ){\mathcal U}_k\Big (z,\frac{y}{2\sqrt{t}}\Big )\Big \}\, .
\end{eqnarray*} In the above, $\xi$, $\zeta$, and $\eta$ are
smooth cutoff functions on $\mathbb{R}_+$ which vanish when the
argument is greater than $\delta$ and which are equal to $1$ if
the argument is less than $\frac12\delta$ where $\delta>0$ is
suitably chosen. We set $\Xi =1-\xi$.  The functions $u_k$ belong
to $C^\infty (M_\Sigma )$,  $U_k^D\in {\mathcal E}(\CD,\kappa )$,
$U_k^R\in {\mathcal E}(\CR,\kappa )$ and ${\mathcal U}_k\in
\Lambda_\kappa^\mu (\Sigma )$ with arbitrary $\mu\in [0,1/2)$.
Moreover the coefficient{  $U_0^R$ equals} zero.
\end{sats}

The asymptotic expansion in this theorem is to be understood in
the sense that the difference \begin{eqnarray}
r_N(x,t)&=&u(x,t)-\sum _{k=0}^N t^{k/2}\Big \{\Xi\Big (\frac{\rho
}{2\sqrt{t}}\Big )\Big (u_k(x)+\rho\eta (\nu /\rho )U_k^D\Big
(x^\prime,\frac{\nu }{2\sqrt{t}}\Big )\nonumber\\ &+&\rho\eta (\nu
/\rho )U_k^R\Big (x^\prime,\frac{\nu }{2\sqrt{t}}\Big )\Big
)+\zeta (\rho ){\mathcal U}_k\Big (z,\frac{y}{2\sqrt{t}}\Big )\Big
\}\label{Hu1z} \end{eqnarray} satisfies the estimate
\begin{equation}\label{EsfSol} |\partial
_t^{\alpha_0}\partial_x^\alpha r_N(x,t)|\leq C
\left\{\begin{array}{ll} t^{(N+1)/2-\alpha_0-|\alpha |/2} &
\mbox{for $t\leq \rho ^2$}\\ t^{(N+1-\mu )/2-\alpha_0}\rho ^{\mu
-|\alpha |} & \mbox{for $t\geq \rho^2$.} \end{array}\right.
\end{equation}  Moreover, the same estimate (in a neighborhood of
$\Sigma $) is valid  for all the derivatives of $r_N$ with respect
to $z$ where the constant involved may depend on the number of
derivatives.

\subsection{The main term} We are looking for $u$ in the form
$$u(x,t)=\textstyle\Xi (\frac{\rho }{2\sqrt{t}} )u_0(x)+v(x,t),$$
where $u_0=\Phi$ and $v$ satisfies the equations: \begin{equation}\label{Ur11b} \left\{\begin{array}{lll} (\partial
_t+D(x,\partial_x))v=f & \mbox{in $M\times (0,T)$},\\ v=g &
\mbox{on $\CD\times (0,T)$,}\\ Rv=h & \mbox{on $\CR\times (0,T)$,} \end{array}\right. \end{equation} with initial
condition \begin{equation}\label{Ur21b} v=0\;\;\;\mbox{on $M$ for $t=0$.} \end{equation} Here
\begin{eqnarray}\label{Ur21bm}
&& \textstyle f(x,t)=-\Xi (\frac{\rho }{2\sqrt{t}}
)Du_0(x)+[\partial_t+D,\xi  (\frac{\rho }{2\sqrt{t}} )]u_0(x),\\
&&\textstyle g=-\Xi (\frac{\rho }{2\sqrt{t}}
)u_0(x),\quad\text{and}\quad h=-R\Xi (\frac{\rho }{2\sqrt{t}}
)u_0(x)\, ,\nonumber \end{eqnarray}
where here and elsewhere $[\cdot,\cdot]$ denotes the commutator of two operators.
Decompose the function $u$
near $\Sigma$ in an asymptotic series with respect to $\rho $: $$
f(x,t)\sim \Xi\Big (\frac{\rho }{2\sqrt{t}}\Big )f_{01}(x)+\zeta
(\rho )\sum_{k=0}^\infty t^{-1+k/2}{\mathcal F}_{0k}\Big
(z,\frac{y}{2\sqrt{t}}\Big )\, , $$ where $f_{01}\in C^\infty
(M_\Sigma )$ and ${\mathcal F}_{0k}(z,Y)$ belongs to
$\Lambda_\kappa^\mu (\Sigma )$ and is equal to zero for small
$|Y|$.  The asymptotic expansion given above means that the
remainder $$ q_N(x,t)=f(x,t)-\Xi\Big (\frac{\rho }{2\sqrt{t}}\Big
)f_{01}(x)-\sum_{k=0}^N t^{-1+k/2}{\mathcal F}_{0k}\Big
(z,\frac{y}{2\sqrt{t}}\Big ) $$
satisfies the estimate
 $$
|\partial_t^{\alpha_0}\partial_x^\alpha \partial_z^\gamma
q_N(x,t)|\leq |C t^{(N-1)/2-\alpha_0-|\alpha |/2} \, ,
$$
where $q_N(x,t)=0$ for $\rho\leq\varepsilon\sqrt{t}$ and for some small positive $\varepsilon$.

Analogously, one can represent $g$  and $h$ as
$$
g(x^\prime,t)=\Xi\Big (\frac{\rho }{2\sqrt{t}}\Big )\rho
g_{01}(x^\prime)+\zeta (\rho )\Xi\Big (\frac{\rho }{2\sqrt{t}}\Big
) g_{00}(z)\, , $$ where $g_{00}\in C^\infty (\Sigma )$  and
$g_{01}\in C^\infty (\CD)$, and $$ h(x^\prime,t)\sim \Xi\Big
(\frac{\rho }{2\sqrt{t}}\Big )h_{01}(x^\prime)+\zeta (\rho
)\rho^{-1}\sum_{k=0}^\infty t^{k/2}{\mathcal H}_{0k}\Big
(z,\frac{\rho}{2\sqrt{t}}\Big )\, , $$ where  $h_{01}\in C^\infty
(\CR)$ and ${\mathcal H}_{0k}(z,Y)$ are smooth functions from
$R_\mu (\Sigma )$ which are equal to $0$ for $|Y|\leq\varepsilon$.

  Now, the function ${\mathcal U}_0$ can be found by solving the problem \begin{eqnarray*} &&\textstyle (\partial_t+L(z,\partial_y)){\mathcal U}_0 (z,\frac{y}{2\sqrt{t}} )=t^{-1} (\Xi (\frac{\rho }{2\sqrt{t}} ) (\frac{\rho}{\sqrt{t}} )^{-2}u_{00}(z,\theta )+{\mathcal F}_{00} (z,\frac{y}{2\sqrt{t}} ) ),\\ &&\textstyle{\mathcal U}_0(z,y_1,0)=\Xi (\frac{\rho }{2\sqrt{t}} )g_{00}(z)\;\;\mbox{for $y_1>0$,}\\ &&\textstyle\rho (R_0{\mathcal U}_0)(z,y_1,0)={\mathcal H}_{00} (z,\frac{\rho }{2\sqrt{t}} )\;\;\mbox{for $y_1<0$.} \end{eqnarray*} By Theorem \ref{T1d}, this boundary value problem has a solution from $\Lambda_\kappa^\mu (\Sigma )$. Similarly, the function $U_0^D$ satisfies the relations $$\textstyle (\partial_t-a\partial^2_\nu )U_0^D (\frac{\nu }{2\sqrt{t}} )=0,\;\;\; U_0^D(0)=
g_{01}(x^\prime)
$$
and, by Proposition \ref{Prop1a}, has a solution $U\in {\mathcal
E}(\CD ,\kappa )$. The remainder $$\textstyle w(x,t)=u(x,t)-\Xi
(\frac{\rho }{2\sqrt{t}} ) (u_0(x)+\eta (\nu /\rho )\rho U_0^D
(x^\prime,\frac{\nu}{2\sqrt{t}} ) )-\zeta (\rho ){\mathcal U}_0
(\frac{y}{2\sqrt{t}} ) $$  satisfies the Equations
\begin{equation}\label{Ur11} \left\{\begin{array}{lll} (\partial
_t+D(x,\partial_x))w=f & \mbox{in $M\times (0,T)$},\\ w=g &
\mbox{on $\CD\times (0,T)$,}\\ Rw=h & \mbox{on $\CR\times (0,T)$,}
\end{array}\right. \end{equation} and the initial condition
\begin{equation}\label{Ur21} w=0\;\;\;\mbox{on $M$ for $t=0$.}
\end{equation} One can verify that the right-hand sides in
(\ref{Ur11}) admit the following asymptotic expansions as $t\to
0$: \begin{eqnarray}\label{Hu11} f(x,t)&\sim& \sum _{k=1}^\infty
t^{k/2-1}\Big \{\Xi\Big (\frac{\rho }{2\sqrt{t}}\Big )\Big
(f_k(x)+\rho\eta (\nu /\rho )F_k^D\Big (x^\prime,\frac{\nu
}{2\sqrt{t}}\Big )\nonumber\\ &+&\rho\eta (\nu /\rho )F_k^R\Big
(x^\prime,\frac{\nu }{2\sqrt{t}}\Big )\Big )+\zeta (\rho
){\mathcal F}_k\Big (z,\frac{y}{2\sqrt{t}}\Big )\Big \}
\end{eqnarray} with $f_k\in C^\infty (M_\Sigma )$, $F_k^D\in
{\mathcal E}(\CD,\varkappa )$, $F_k^R\in {\mathcal
E}(\CR,\varkappa )$ and ${\mathcal F}_k\in
\Lambda_\varkappa^{\mu -1}(\Sigma )$,
\begin{equation}\label{Hu11a} g(x^\prime,t)\sim \sum _{k=1}^\infty
t^{k/2}\Big (\Xi\Big (\frac{\rho }{2\sqrt{t}}\Big )\rho
g_k(x^\prime)+\zeta (\rho ){\mathcal G}_k\Big (z,\frac{\rho
}{2\sqrt{t}}\Big )\Big )\, , \end{equation} with $g_k\in C^\infty
(\CD)$, ${\mathcal G}_k\in {\mathcal R}_{\mu +1}(\Sigma)$, and
\begin{equation}\label{Hu11b} h(x^\prime,t)\sim \sum _{k=1}^\infty
t^{(k-1)/2}\Big (\Xi\Big (\frac{\rho }{2\sqrt{t}}\Big
)h_k(x^\prime)+\zeta (\rho )\frac{\sqrt{t}}{\rho}{\mathcal
H}_k\Big (z,\frac{\rho }{2\sqrt{t}}\Big )\Big ) \end{equation}
with $h_k\in C^\infty (\CR)$, ${\mathcal H}_k\in {\mathcal R}_{\mu
+1}(\Sigma)$. The asymptotic expansion of Equation (\ref{Hu11}) is
to be understood in the following sense. Let \begin{eqnarray*}
&&f^{(N)}(x,t)= \sum _{k=1}^N t^{k/2-1}\Big \{\Xi\Big (\frac{\rho
}{2\sqrt{t}}\Big )\Big (f_k(x)+\rho\eta (\nu /\rho )F_k^D\Big
(x^\prime,\frac{\nu }{2\sqrt{t}}\Big )\nonumber\\ &&+\rho\eta (\nu
/\rho )F_k^R\Big (x^\prime,\frac{\nu }{2\sqrt{t}}\Big )\Big
)+\zeta (\rho ){\mathcal F}_k\Big (z,\frac{y}{2\sqrt{t}}\Big
)\Big \} \end{eqnarray*} then the remainder $R_N=f-f^{(N)}$ satisfies
the estimate \begin{equation}\label{RemEs1a}
|\partial_t^{\alpha_0}\partial_x^\alpha \partial_z^\gamma
R_N(x,t)|\leq C \left\{\begin{array}{ll}
t^{(N-1)/2-\alpha_0-|\alpha |/2} & \mbox{for $t\leq \rho ^2$}\\
t^{(N-1)/2-\alpha_0+(1-\mu )/2}\rho^{\mu -1-|\alpha|} & \mbox{for
$t\geq \rho^2$,} \end{array}\right. \end{equation} for all
$\alpha_0=0,1,\ldots$ and multi-indices $\alpha =(\alpha _1,\cdot
,\alpha_m)$ and $\gamma =(\gamma_1,\ldots ,\gamma_{m-2})$ with
nonnegative integer components. The derivatives with respect to
$z$ are defined and should be taken into account only in a
neighborhood of $\Sigma $, outside of this neighborhood $\gamma$
is zero.

If we denote by $g^{(N)}(x^\prime,t)$ and $h^{(N)}(x^\prime,t)$
the partial sums in (\ref{Hu11a}) and (\ref{Hu11b}) from $1$ to
$N$ and introduce the remainder terms $R_{gN}=g-g^{(N)}$ and
$R_{hN}=h-h^{(N)}$ then the asymptotic representations
(\ref{Hu11a}) and (\ref{Hu11b}) mean that
\begin{equation}\label{RemEs1b}
|\partial_t^{\alpha_0}\partial_{x^\prime}^{\alpha
'}\partial_z^\gamma R_{gN}(x,t)|\leq C \left\{\begin{array}{ll}
t^{(N+1)/2-\alpha_0-|\alpha |/2} & \mbox{for $t\leq \rho ^2$}\\
t^{N/2-\alpha_0-\mu /2}\rho^{\mu +1-|\alpha|} & \mbox{for $t\geq
\rho^2$,} \end{array}\right. \end{equation} and
\begin{equation}\label{RemEs1c}
|\partial_t^{\alpha_0}\partial_{x^\prime}^{\alpha
'}\partial_z^\gamma R_{hN}(x,t)|\leq C \left\{\begin{array}{ll}
t^{N/2-\alpha_0-|\alpha |/2} & \mbox{for $t\leq \rho ^2$}\\
t^{N/2-\alpha_0-\mu /2}\rho^{\mu -|\alpha|} & \mbox{for $t\geq
\rho^2$,} \end{array}\right. \end{equation} for all
$\alpha_0=0,1,\ldots$ and multi-indices $\alpha '=(\alpha_1,\cdots
,\alpha _{m-1})$ and $\gamma =(\gamma_1,\ldots ,\gamma_{m-2})$.
Roughly speaking the estimates for the remainder terms are the
same as the estimate for the next terms in the asymptotic
expansions (\ref{Hu11})--(\ref{Hu11b}).

The form of the right-hand sides (\ref{Hu11})--(\ref{Hu11b}) are
more general than we need but it is convenient to consider this
more general form in order to unify the construction of other
terms in the asymptotic expansion for $u$.

\subsection{Higher order terms}\label{SubZ}
We first describe the construction of the terms $u_1$, $U_1^D$,
$U_1^R$ and ${\mathcal U}_1$.

{\em The term $u_1$.} We take $u_1= 2f_1$. Then \begin{eqnarray*}
&&(\partial _t+D(x,\partial_x))\Big (t^{1/2}\Xi\Big (\frac{\rho
}{2\sqrt{t}}\Big )u_1(x)\Big )-t^{-1/2}\Xi\Big (\frac{\rho
}{2\sqrt{t}}\Big )f_1(x)\nonumber\\ &&=\frac{\rho }{4t}\xi '\Big
(\frac{\rho }{2\sqrt{t}}\Big )u_1(x)-t^{1/2}\Big
[D(x,\partial_x),\xi\Big (\frac{\rho }{2\sqrt{t}}\Big )\Big
]u_1(x)\nonumber\\ &&+t^{1/2}\Xi\Big (\frac{\rho }{2\sqrt{t}}\Big
)D(x,\partial_x)u_1(x)\,. \end{eqnarray*} Decomposing the function
$u_1$ near $\Sigma$ in asymptotic series with respect to $\rho$ we
obtain that the right-hand side is asymptotically equal to $$
t^{1/2}\Xi\Big (\frac{\rho }{2\sqrt{t}}\Big
)f_{11}(x)+\sum_{k=1}^\infty t^{k/2-1}\zeta (\rho ){\mathcal
F}_{1k}\Big (z,\frac{y}{2\sqrt{t}}\Big ) $$ with $f_{11}\in
C^\infty (M_\Sigma )$ and ${\mathcal F}_{1k}\in\Lambda _\kappa^\mu
(\Sigma )$. Moreover ${\mathcal F}_{1k}(z,Y)=0$ for
$|Y|\leq\varepsilon$ for some positive $\varepsilon$.  So, we have compensated the term containing
$f_1$ in the right-hand side of (\ref{Hu11}) and the discrepancy,
which came, can be included in the remaining terms in the
right-hand side in (\ref{Hu11}). We shall denote the new
right-hand sides by the same letters.

\bigskip
{\em The term ${\mathcal U}_1$.} We find this function from the
equation $$ (\partial_t-L(z,\partial_y))\Big (t^{1/2}{\mathcal
U}_1\Big (z,\frac{y}{2\sqrt{t}}\Big )\Big )=t^{-1/2}{\mathcal
F}_1\Big (z,\frac{y}{2\sqrt{t}}\Big ) $$ supplied with boundary
conditions $$ {\mathcal U}_1(z,y_1,0)={\mathcal
G}_1(z,y_1)\;\;\;\mbox{for $y_1>0$} $$ and $$
\rho\sum_{j=1}^2A^{j2}(z)(\partial_{y_j}{\mathcal
U}_1)(z,y_1,0)={\mathcal H}_1(z,-y_1)\;\;\;\mbox{for $y_1<0$.} $$
The discrepancy in the equation brought by this term is equal to
\begin{eqnarray*} &&(\partial_t+D)\Big (t^{1/2}\zeta (\rho
){\mathcal U}_k\Big (z,\frac{y}{2\sqrt{t}}\Big )\Big
)-t^{-1/2}\zeta (\rho ){\mathcal F}_k\Big
(z,\frac{y}{2\sqrt{t}}\Big )\nonumber\\ &&=t^{1/2}\zeta (\rho
)(D+L){\mathcal U}_k\Big (z,\frac{y}{2\sqrt{t}}\Big
)+t^{1/2}[D,\zeta (\rho)]{\mathcal U}_k\Big
(z,\frac{y}{2\sqrt{t}}\Big ). \end{eqnarray*} Using (\ref{Rep2})
and the asymptotic expansion at infinity for functions from the
class $\Lambda $ one can show that the right-hand side of the
equation above has asymptotics
\begin{eqnarray*}
&&\sum _{k=2}^\infty t^{k/2-1}\Big \{\Xi\Big (\frac{\rho
}{2\sqrt{t}}\Big )\Big (f_{1k}(x)+\eta (\nu /r )F_{1k}^D\Big
(x^\prime,\frac{\nu }{2\sqrt{t}}\Big )\nonumber\\ &&+\eta (\nu /r
)F_{1k}^R\Big (x^\prime,\frac{\nu }{2\sqrt{t}}\Big )\Big )+\zeta
(\rho ){\mathcal F}_{2k}\Big (z,\frac{y}{2\sqrt{t}}\Big )\Big \}\,
, \end{eqnarray*} where $f_{1k}\in C^\infty (M_\Sigma )$,
$F_{1k}^D\in {\mathcal E}(\CD,\kappa )$, $F_{1k}^R\in {\mathcal
E}(\CR,\kappa )$ and ${\mathcal F}_{2k}\in\Lambda_\kappa^{\mu
-1}(\Sigma )$.

The discrepancy in the Dirichlet boundary condition is zero and in
the Robin boundary condition is \begin{eqnarray*} &&R(t^{1/2}\zeta
{\mathcal U}_1)-\zeta (\rho )\rho ^{-1}{\mathcal H}_{k}\\ &&=\sum
_{k=2}^\infty t^{(k-1)/2}\Big (\Xi\Big (\frac{\rho
}{2\sqrt{t}}\Big )h_{1k}(x^\prime)+\zeta (\rho
)\frac{\sqrt{t}}{\rho}{\mathcal H}_{1k}\Big (z,\frac{\rho
}{2\sqrt{t}}\Big )\Big )\, , \end{eqnarray*} where $h_{1k}\in
C^\infty (\CR)$, ${\mathcal H}_{1k}\in {\mathcal R}_{\mu
+1}(\Sigma)$. So, one can see that ${\mathcal U}_1$ compensates
the terms ${\mathcal F}_1$, ${\mathcal G}_1$ and ${\mathcal H}_1$
in the right-hand sides of (\ref{Hu11})--(\ref{Hu11b}). The
discrepancies brought by ${\mathcal U}_1$ have lower order and can
be included in terms in the asymptotic expansions
 (\ref{Hu11})--(\ref{Hu11b}) with $k\geq 2$.

\bigskip
{\em The terms $U_1^D$ and $U_1^R$.} We define the function $U_1^D$, $x^\prime\in \CD$, as a solution of the boundary value problem \begin{eqnarray*} &&(\partial_t -a(x^\prime)\partial_\nu^2)(t^{1/2}U_1^D(x^\prime,\nu ))=t^{-1/2}F_1^D(x^\prime,\nu ),\\
&&U_1^D(x^\prime,0)=g_1(x^\prime)
\end{eqnarray*}
with $x^\prime$ considered as a parameter. We have
\begin{eqnarray*} &&(\partial_t+D(x,\partial_x))\Big
(t^{1/2}\eta\Big (\frac{\nu }{\rho} \Big )\rho
U_1^D\Big (x^\prime,\frac{\nu}{2\sqrt{t}}\Big )\Big )\\&&\qquad\qquad-t^{-1/2}\Xi \Big (\frac{\rho }{2\sqrt{t}}\Big )\eta\Big (\frac{\nu }{\rho}\Big )\rho F_1^D\Big (x^\prime,\frac{\nu}{2\sqrt{t}}\Big )\nonumber\\
&&=t^{1/2}\Xi\Big (\frac{\rho }{2\sqrt{t}}\Big )\eta\Big
(\frac{\nu }{\rho}\Big
)(D(x,\partial_x)+a(x^\prime)\partial_\nu^2)\rho U_1^D\Big
(x^\prime,\frac{\nu}{2\sqrt{t}}\Big )\nonumber\\  &&+t^{1/2}\Big
[D(x,\partial_x),\Xi\Big (\frac{\rho }{2\sqrt{t}}\Big )\eta\Big
(\frac{\nu }{\rho}\Big )\Big ]\rho U_1^D\Big
(x^\prime,\frac{\nu}{2\sqrt{t}}\Big )\, .\nonumber \end{eqnarray*}
By Equation (\ref{Rep1}), the right-hand side in this equation
asymptotically equals $$ \sum_{j=2}^\infty t^{j/2-1}\zeta (\rho
){\mathcal F}_{1j}\Big (z,\frac{y}{2\sqrt{t}}\Big )
+\rho\sum_{j=2}^\infty t^{j /2-1}\Xi\Big (\frac{\rho
}{2\sqrt{t}}\Big )\eta (\nu /\rho )F_{1j}^D\Big
(x^\prime,\frac{\nu}{2\sqrt{t}}\Big ) $$ with ${\mathcal
F}_{1j}\in \Lambda _\kappa^{\mu -1} (\Sigma )$ and $F_{1j}^D\in
{\mathcal E}(\CD,\kappa )$.

The term $U_1^R$ is then constructed analogously. Thus, we have
constructed the approximation of the solution which contains all
terms in the asymptotic expansion of Theorem \ref{TQuQu} with
$k=1$ and which compensates all terms in the asymptotic
representations (\ref{Hu11})--(\ref{Hu11b}) of the right-hand
sides of (\ref{Ur11}) with $k=1$. Continuing this procedure we can
compensate terms with $k=2,3,\dots$. Therefore, if we put
\begin{eqnarray*} &&u_N(x,t)=\sum _{k=0}^N t^{k/2}\Big \{\Xi\Big
(\frac{\rho }{2\sqrt{t}}\Big )\Big (u_k(x)+\rho\eta\Big (\frac{\nu
}{\rho } \Big )U_k^D\Big (x^\prime,\frac{\nu }{2\sqrt{t}}\Big
)\nonumber\\ &&+\rho\eta\Big (\frac{\nu }{\rho}\Big )U_k^R\Big
(x^\prime,\frac{\nu }{2\sqrt{t}}\Big )\Big )+\zeta (\rho
){\mathcal U}_k\Big (z,\frac{y}{2\sqrt{t}}\Big )\Big \}
\end{eqnarray*} and $r_N=u-u_N$. The function $r_N$ then satisfies
the equations: $$ \left\{\begin{array}{lll} (\partial
_t+D(x,\partial_x))r_N=f_N & \mbox{in $M\times (0,T)$}\\ r_N=g_N &
\mbox{on $\CD\times (0,T)$,}\\ Ru_N=h_N & \mbox{on $\CR\times
(0,T)$,} \end{array}\right. $$ with the initial condition $
r_N=0\;\;\;\mbox{on $M$ for $t=0$}$. The right-hand sides in these
relations admit the asymptotic expansions
(\ref{Hu11})--(\ref{Hu11b}), respectively, where summation is
started from $k=N+1$.

\section{Estimate of the remainder term in the asymptotics}\label{sect-8} In this section, we complete the proof of Theorem \ref{TQuQu}
 by establishing the remainder estimate given in Equation \ref{EsfSol}. The proof rests on a result (Theorem \ref{ThX} below)
obtained by Johansson \cite{J}. First we introduce some additional notation. Let $T$ be a positive number and $Q_T=M\times (0,T)$,
$\Gamma_T^D=\CD\times (0,T)$ and $\Gamma_T^R=\CR\times (0,T)$. We introduce also weighted Sobolev spaces. Let $\ell =0,1,\ldots
$, $\beta\in\mathbb{R}$. The space $W^{2\ell ,\ell }_\beta (Q_T)$ consists of functions on $Q_T$ with the finite norm $$
||u||_{W^{2\ell ,\ell }_\beta (Q_T)}=\Big (\int_{Q_T}\rho^{2(\beta -2\ell )}\sum_{|\overline{\alpha }|\leq 2\ell
}\rho^{2|\overline{\alpha}|}|\partial_t^{\alpha_0}\partial_x^\alpha u|^2dxdt\Big )^{1/2} $$ where we set $\overline{\alpha
}:=(\alpha_0,\alpha )$ and
$|\overline{\alpha}|:=2\alpha_0+|\alpha|$. For $s=1/4,3/4,\ldots
$, introduce the trace spaces $W^{2s,s}_\beta (\Gamma _T^D)$ with
the norm $$\textstyle ||u||_{W^{2s,s}_\beta (\Gamma _T^D)}=
(\int_0^T||u(\cdot,t)||_{V^{2s}_\beta
(\CD)}^2dt+\int_{\CD}\rho^{2\beta}||u(x,\cdot )||_{H^s(0,T)}^2dx
)^{1/2} \,.$$ Here $H^s$ stands for the standard Sobolev space on the
interval $(0,T)$. If $k$ is a positive integer, then
$V^{k-1/2}_\beta (\CD)$ is the space of traces on $\CD$ of
functions from the space $V^k_\beta (M)$ with the norm
$$\textstyle ||v||_{V^k_\beta (M)}= (\int_M\sum_{|\alpha |\leq
k}\rho^{2(\beta-k+|\alpha |)}|\partial_x^\alpha v(x)|^2dx
)^{1/2}\, .
$$
The norm in $V^{k-1/2}_\beta (\CD)$ is defined by
$$
||w||_{V^{k-1/2}_\beta (\CD)}=\inf \{\, ||v||_{V^k_\beta (M)}\,
:\, v\in V^k_\beta (M), v\big |_{\CD}=w\}\, .
$$
Analogously, one can define the space $W^{2s,s}_\beta (\Gamma
_T^D)$.

The closure of functions from the space $W^{2\ell ,\ell }_\beta
(Q_T)$ equal to $0$ for small $t$ will be denoted by $W^{2\ell
,\ell }_{\beta ,0}(Q_T)$. Analogously one defines the spaces
$W^{2s,s}_{\beta ,0}(\Gamma _T^D)$ and $W^{2s,s}_{\beta ,0}(\Gamma
_T^R)$.

If $|\overline{\alpha}|<2\ell -n/2-1$ then functions $u\in
W^{2\ell ,\ell }_\beta (Q_T)$ have continuous derivatives of order
$\overline{\alpha}$ in $Q_T$ and $$ |\partial
_t^{\alpha_0}\partial_x^\alpha u(x,t)|\leq C\rho^{2\ell
|-|\overline{\alpha }|-n/2-1-\beta}\, ||u||_{W^{2\ell ,\ell
}_\beta (Q_T)}\, .
$$
This estimate can be obtained from the analogous estimate for
functions from nonweighted spaces (see \cite{BIN}, Chapter 3) and
homogeneity arguments. If $u\in W^{2\ell ,\ell }_{\beta ,0}(Q_T)$
and $|\overline{\alpha}|+2m<2\ell -n/2-1$ with a nonnegative $m$
then, clearly, \begin{equation}\label{EstCont} |\partial
_t^{\alpha_0}\partial_x^\alpha u(x,t)|\leq Ct^m\rho^{2\ell
|-|\overline{\alpha }|-n/2-1-\beta}\, ||u||_{W^{2\ell ,\ell
}_\beta (Q_T)}\, . \end{equation}

The proof of the following result is contained in \cite{J}.

\begin{sats}\label{ThX} Let $\ell \geq 1$ be an integer and let $\beta $ satisfy $1/2 <-\beta +2\ell <3/2$. If $f\in W^{2\ell -2,\ell -1}_{\beta ,0}(Q_T)$, $g\in W^{2\ell -1/2,\ell -1/4}_{\beta ,0}(\Gamma _T^D)$ and $h\in W^{2\ell -3/2,\ell -3/4}_{\beta ,0}(\Gamma _T^R)$ then there exists a unique solution $u\in W^{2\ell ,\ell }_{\beta ,0}(Q_T)$ to problem {\rm (\ref{Ur11})}.
This solution satisfies the estimate
\begin{eqnarray}\label{Est1a} ||u||_{W^{2\ell ,\ell }_\beta (Q_T)}&\leq& C\Big
(||f||_{W^{2\ell -2,\ell -1}_\beta  (Q_T)}+||g||_{W^{2\ell -1/2,\ell -1/4}_\beta (\Gamma _T^D)}\nonumber\\
&&\qquad+||h||_{W^{2\ell -3/2,\ell -3/4}_\beta (\Gamma _T^R)}\Big ). \end{eqnarray}

\end{sats}

Now we are in a position to prove the remainder estimate
(\ref{EsfSol}). According to the construction of the terms in the
asymptotic expansion given in Theorem \ref{TQuQu} (see the end of
Sect. \ref{SubZ}) the remainder (\ref{Hu1z}) satisfies the
boundary value problem (\ref{Ur11}), (\ref{Ur21}), where the
right-hand sides admit the asymptotic representations
(\ref{Hu11})--(\ref{Hu11b}) with summation starting with $k=N+1$.
Therefore these right-hand sides are estimated by the right--hand
sides in (\ref{RemEs1a})--(\ref{RemEs1c}). This implies that
the derivative of order $k$ with respect to $t$  and all
derivatives with respect to $z$ (in a neighborhood of $\Sigma$)
belong to
$$
W^{2\ell -2,\ell -1}_{\beta ,0}(Q_T)\, ,\;\;\; W^{2\ell -1/2,\ell
-1/4}_{\beta ,0}(\Gamma _T^D)\;\;\; \mbox{and}\;\;\; W^{2\ell
-3/2,\ell -3/4}_{\beta ,0}(\Gamma _T^R), $$ for $\ell <(N-2k-1)/2$
and $2\ell -\beta <2+\mu $, respectively. We suppose here  that
$\mu $ is an arbitrary number from the interval $(0,1/2)$. Now
applying Theorem \ref{ThX} we obtain that $\partial_t^kr_N$
together with  all derivatives with respect to $z$ (in a
neighborhood of $\Sigma$) belongs to $W^{2\ell ,\ell }_{\beta
,0}(Q_T)$ for $1/2 <-\beta +2\ell <1+\mu $ and $\ell <(N-2k-1)/2$.
This implies that in a +neighborhood of $\Sigma$ the integral
$$
\textstyle\int\rho^{2(\beta -2\ell)}\sum_{|\alpha |\leq
2\ell}\rho^{2|\alpha|}|\partial_y^\alpha\partial_t^{k-1}\partial_z^\gamma
u|^2dy $$ is bounded uniformly with respect to $t$ and $z$. By the
usual imbedding theorem we obtain $$
|\partial_y^\alpha\partial_t^{k-1}\partial_z^\gamma u|\leq C\rho
^{2\ell -|\alpha|-1-\beta}\,
$$
for $|\alpha |< 2\ell -1$. Choosing $\sigma =2\ell -1-\beta$ close
to $1/2$ and then taking $\mu\in (\sigma ,1/2)$ we can rewrite the
above estimate as \begin{equation}\label{Rem2m}
|\partial_y^\alpha\partial_t^{k-1}\partial_z^\gamma u|\leq
Ct^m\rho ^{\sigma -|\alpha|}\,
\end{equation}
which is valid for $|\alpha|+2k+2m <N-2$ and for arbitrary
multi-index $\gamma$.

In order to obtain a remainder estimate outside a neighborhood of
$\Sigma$ one can use (\ref{EstCont}) which gives
\begin{equation}\label{Rem2}
|\partial _t^{\alpha_0}\partial_x^\alpha r_N(x,t)|\leq Ct^m
\end{equation}
for $|\overline{\alpha}|+2m<2\ell -n/2-1$ and for
$\rho\geq\varepsilon$ where $\varepsilon$ is a small positive
number. Estimates (\ref{Rem2m}) and (\ref{Rem2}) imply
\begin{equation}\label{Rem2w} |\partial
_t^{\alpha_0}\partial_x^\alpha r_N(x,t)|\leq Ct^m\rho ^{\sigma
|-|\alpha|}
\end{equation}
for $|\overline{\alpha}|+2m<N -n/2-2$ and for arbitrary $\sigma\in
(0,1/2)$.

In order to obtain estimate (\ref{EsfSol}) for $r_N$ we proceed as
follows. We choose an integer $M>N$ and represent the remainder
term $r_N$ as \begin{eqnarray}\label{Rem1s} &&r_N=r_M+\sum
_{k=N+1}^M t^{k/2}\Big \{ \Xi\Big (\frac{\rho }{2\sqrt{t}}\Big
)\Big  (u_k(x)+\rho\eta (\nu /\rho )U_k^D\Big (x^\prime,\frac{\nu
}{2\sqrt{t}}\Big )\nonumber\\ &&+\rho\eta (\nu /\rho )U_k^R\Big
(x^\prime,\frac{\nu }{2\sqrt{t}}\Big )\Big ) +\zeta (\rho
){\mathcal U}_k\Big (z,\frac{y}{2\sqrt{t}}\Big )\Big \} .
\end{eqnarray}
One can check that all the terms in the summation satisfy estimate
(\ref{EsfSol}).
By choosing $M$ sufficiently large, we obtain estimate
(\ref{EsfSol}) for $r_M$ from  (\ref{Rem2w}). The proof of Theorem \ref{TQuQu} is complete.\hfill\qedbox

\medbreak Let $\Sigma_D$ and $\Sigma_R$ denote the boundaries of $C_D$ and $C_R$, respectively. Clearly, functions from $C^\infty
(M_\Sigma )$ may take different values on $\Sigma_D$ and
$\Sigma_R$. The  existence of the asymptotic series given in Theorem \ref{thm-1.1} is a special case of the following more
general result:

\begin{theorem}\label{thm-8.2}
Let $u$ be the solution to problem (\ref{eqn-1.c}) and let
$\sigma\in C^\infty (M_\Sigma )$. Then the following asymptotic
expansion for $u$ is valid:
\begin{equation}\label{Linkop1}
\int_M\sigma (x)u(x,t)dx \sim\sum_{k=0}^\infty t^k\int_M\sigma (x)
u_k(x)dx+\sum_{k=0}^\infty t^{k/2}\Big (ta_k
+t^{1/2}b_k+t^{3/2}c_k\Big ),
\end{equation}
where
$$
a_k=\int_\Sigma\int_{-\pi}^{\pi}v_k(z,\theta )dzd\theta\, ,\;\;\;
b_k=\int_{C_D}w_k^D(x')dx'+\int_{C_R}w_k^R(x')dx'
$$
and
$$
c_k=\int_{\Sigma_D} h_k^D(z)dz+\int_{\Sigma_R} h_k^R(z)dz\, .
$$
Here $u_k$, $v_k$, $w^D_k$, $w^R_k$, $h_k^D$ and $h_k^R$ are
smooth functions whose values at a given point depend only on values of
$\varphi $ and its derivatives at this point.
\end{theorem}

\begin{proof} We have
\begin{equation}\label{Linkop1a}
\int_M\sigma (x)\Xi\Big (\frac{\rho}{2\sqrt{t}}\Big
)u_k(x)dx=\int_M\sigma (x)u_k(x)dx-\int_M\xi\Big
(\frac{\rho}{2\sqrt{t}}\Big )\sigma (x)u_k(x)dx\, .
\end{equation}
From $\sigma ,\, u_k\in C^\infty (M_\Sigma )$ it follows that
$$
\sigma (x)u_k(x)=\sum_{j=0}^Nu_{kj}(z,\theta )\rho^j+O(\rho^{N+1})
$$
for each $N$, which implies that
$$
\int_M\xi\Big (\frac{\rho}{2\sqrt{t}}\Big )\sigma
(x)u_k(x)dx=t\sum_{j=0}^\infty
c_j\int_\Sigma\int_{-\pi}^{\pi}u_{kj}(z,\theta )dzd\theta\,
t^{j/2},
$$
where $c_j$ are constants independent of the initial data
$\varphi$. Therefore, the integrals in (\ref{Linkop1a}) give the
first sum  in the right-hand side of (\ref{Linkop1}) in  Theorem \ref{thm-8.2} and terms of
the form $ta_k$ in (\ref{Linkop1}).

Next, consider the integral
\begin{equation}\label{Linkop2}
\int_M\sigma (x)\Xi\Big (\frac{\rho}{2\sqrt{t}}\Big )\rho\eta\Big
(\frac{\nu}{ \rho}\Big )U_k^D\Big (x',\frac{\nu }{2\sqrt{t}}\Big
)dx\,.
\end{equation}
Let us introduce a cutoff function $\zeta (y_1)$ which is equal to
$1$ for $|y_1|\leq\varepsilon /2$ and $0$ for $|y_1|\geq
\varepsilon$, where $\varepsilon$ is a small positive number. Then
we represent (\ref{Linkop2}) for small $t$ as
\begin{eqnarray}\label{Linkop3}
&&\!\int_M \!(1-\zeta\! (y_1))\rho\eta\Big (\frac{\nu}{\rho}\Big
)\sigma (x)U_k^D\Big (x',\frac{\nu }{2\sqrt{t}}\Big
)dx\nonumber\\
&&+\int_\Sigma\!\int_0^\infty\!\!\int_0^\infty\!\!\!\zeta
(y_1)\rho \sigma (z,y)U_k^D\!\Big
(z,y_1,\frac{y_2 }{2\sqrt{t}}\!\Big )\!dydz\nonumber\\
&&+\int_\Sigma\int_0^\infty\int_0^\infty\zeta (y_1)\rho (\eta\Big
(\frac{y_2}{\rho}\Big )-1)\sigma (z,y)U_k^D\Big
(z,y_1,\frac{y_2 }{2\sqrt{t}}\Big )dydz\nonumber\\
&&-\int_\Sigma\int_0^\infty\int_0^\infty\xi\Big
(\frac{\rho}{2\sqrt{t}}\Big )\rho\eta\Big (\frac{y_2}{\rho}\Big
)\sigma (z,y)U_k^D\Big (z,y_1,\frac{y_2 }{2\sqrt{t}}\Big )dy dz
\end{eqnarray}
where $\rho$ is equal to $\sqrt{y_1^2+y_2^2}$ in coordinates
$y=(y_1,y_2)$. One can check directly that the first two integrals
in (\ref{Linkop3}) have asymptotics
$$
\sum_{j=0}^\infty t^{(1+j)/2}\int_{C_D}q_{kj}(x')dx'
$$
and the third integral has an expansion of the form
$$
\sum_{j=0}^\infty t^{1+j/2}\int_{\Sigma_D} h^{(1)}_{kj}(z)dz\, .
$$
Making change of variables $y_1=2\sqrt{t}Y_1$ and
$y_2=2\sqrt{t}Y_2$ we can rewrite the last integral in
(\ref{Linkop3}) as
$$
4t^{3/2}\int_\Sigma\int_0^\infty\int_0^\infty \xi (\rho )\rho\eta
(Y_2 /\rho )\sigma (z,\sqrt{t}\,Y)U_k^D (z,2\sqrt{t}\,Y_1,Y_2
)dY_1dY_2dz\, .
$$
Since $U_k^D\in {\cal E}(C_D,\kappa )$, the last integral has the
asymptotics
$$
\sum_{j=0}^\infty t^{(3+j)/2}\int_{\Sigma_D} h^{(2)}_{kj}(z)dz\, ,
$$
where $h_{kj}^{(2)}$ are integrals with respect to $Y_2$ of linear
combinations of functions $\partial_{Y_1}^sU_k^D (z,Y_1,Y_2
)|_{Y_1=0}$ multiplied by explicit weights.
 The term
$$
\int_M\sigma (x)\Xi\Big (\frac{\rho}{2\sqrt{t}}\Big )\rho\eta (\nu
/\rho )U_k^R\Big (x',\frac{\nu }{2\sqrt{t}}\Big )dx
$$
is considered analogously.

It remains to obtain an asymptotic expansion of the term
\begin{eqnarray}\label{Linkop4}
\int_\Sigma \int_0^\infty\int_0^\infty\sigma (x)\zeta (\rho ){\cal
U}_k\Big (z,\frac{y}{2\sqrt{t}}\Big )dzdy_1dy_2\, .
\end{eqnarray}
Using the asymptotic expansion for the function ${\cal U}_k$ for
large second argument:
$$\textstyle
{\mathcal U}_k(z,Y)\sim \sum_{j=0}^\infty
|Y|^{-j}\{\frac{v_{kj}(z,\theta )}{|Y|}+\sum_{\pm }U_{kj}^{\pm
}(z,Y_2)\chi (\textstyle\frac{Y_2}{|Y|} ) \}\, ,
$$
we obtain integrals similar to the ones just considered. Moreover, one can
show that the coefficient $v_{00}$ is equal to zero because of
vanishing of the analogous coefficient in the asymptotics of the
right-hand side in the equation for the function ${\mathcal U}_0$.
Reasoning as above we arrive at the required asymptotic
representation for these integrals. This completes the proof of Theorem \ref{thm-8.2} and thereby of Theorem \ref{thm-1.1}.
\end{proof}

\section*{Acknowledgments}
The research of P. Gilkey was partially supported by the Max
Planck Institute for Mathematics in the Sciences (Leipzig,
Germany) and the Institut Mittag-Leffler (Stockholm, Sweden). The
research of K. Kirsten was partially supported by the Max Planck
Institute for Mathematics in the Sciences (Leipzig, Germany) and
by the Baylor University Summer Sabbatical Program. The research
of M. van den Berg was supported by the London Mathematical
Society under Scheme 4 references 4817 and 4407, and by the
Institut Mittag-Leffler (Stockholm, Sweden). The research of V.
Kozlov was supported by the Swedish Research Council.

\end{document}